\documentclass{article}

\usepackage{arxiv}

\usepackage[utf8]{inputenc} 
\usepackage[T1]{fontenc}    
\usepackage{hyperref}       
\usepackage{url}            
\usepackage{booktabs}       
\usepackage{amsfonts}       
\usepackage{nicefrac}       
\usepackage{microtype}      
\usepackage{lipsum}		
\usepackage{graphicx}
\usepackage{natbib}
\usepackage{doi}
\usepackage{float}
\usepackage{enumerate}

\usepackage{amsmath}
\usepackage{bbm}
\usepackage[table]{xcolor}

\theoremstyle{thmstyleone}%
\newtheorem{theorem}{Theorem}
\theoremstyle{thmstyletwo}%
\newtheorem{assumption}{Assumption}%
\theoremstyle{thmstylethree}%

\newcounter{assumptioncounter}

\usepackage[figuresright]{rotating}
\usepackage{amsmath}
\usepackage{amssymb}
\usepackage{bbm}
\usepackage{titlesec}
\usepackage{caption}

\title{Multiply Robust Difference-in-Differences Estimation of Causal Effect Curves for Continuous Exposures}

\author{Gary Hettinger \\
	Department of Biostatistics, Epidemiology, and Informatics \\
	University of Pennsylvania\\
	Philadelphia, PA, U.S.A. \\
	\texttt{ghetting@pennmedicine.upenn.edu} \\
        \And
	Youjin Lee \\
	Department of Biostatistics\\
	Brown University\\
	Providence, RI, U.S.A. \\
        \And
	Nandita Mitra \\
	Department of Biostatistics, Epidemiology, and Informatics \\
	University of Pennsylvania\\
	Philadelphia, PA U.S.A. \\
}

\renewcommand{\shorttitle}{Robust Difference-in-Differences Causal Effect Curves}

\hypersetup{
pdftitle={Multiply Robust Difference-in-Differences Estimation of Causal Effect Curves for Continuous Exposures},
pdfauthor={Gary~Hettinger, Youjin~Lee, Nandita~Mitra},
pdfkeywords={Dose Response, Health Policy, Influence Function, Semi-parametric},
}

\begin{document}
\maketitle

\begin{abstract}
Researchers commonly use difference-in-differences (DiD) designs to evaluate public policy interventions. While methods exist for estimating effects in the context of binary interventions, policies often result in varied exposures across regions implementing the policy. Yet, existing approaches for incorporating continuous exposures face substantial limitations in addressing confounding variables associated with intervention status, exposure levels, and outcome trends. These limitations significantly constrain policymakers’ ability to fully comprehend policy impacts and design future interventions. In this work, we propose new estimators for causal effect curves within the DiD framework, accounting for multiple sources of confounding. Our approach accommodates misspecification of a subset of treatment, exposure, and outcome models while avoiding any parametric assumptions on the effect curve. We present the statistical properties of the proposed methods and illustrate their application through simulations and a study investigating the heterogeneous effects of a nutritional excise tax under different levels of accessibility to cross-border shopping.
\end{abstract}


\section{Introduction}
\label{sec:intro}

Public policies wield significant influence on a broad spectrum of population outcomes, including those related to public health and economics \citep{PollackPorter2018TheProblems}.
Accordingly, decision-makers require rigorous evaluation of such policies to comprehend their effectiveness and inform the design of subsequent initiatives. 
Complicating the comprehensive assessment of policy effects, units within a study may be differentially exposed to a policy for various reasons. 
For example, exposure to an excise tax may depend on the ease in which a buyer can evade the tax (e.g., geographic proximity to cross-border shopping) or responsive price adjustments made by the seller which may diminish the actual tax rate, or pass-through, seen by consumers \citep{Chaloupka2019TheConsumption}.
Such variation can lead to exposure-induced effect heterogeneity that, if understood and isolated from confounder-induced effect heterogeneity, can help identify policy mechanisms and inform future interventions designed to generate a different exposure distribution \citep{Cintron2022HeterogeneousSciences}.

When evaluating effects of policy interventions, researchers commonly use difference-in-differences (DiD) approaches, which exploit the so-called counterfactual parallel trends assumption to utilize outcomes from a control group, measured before and after the intervention, to impute what would have happened in the treated group in the absence of intervention \citep{Heckman1997MatchingProgramme}.
This assumption requires that there are no confounding differences between treated and control groups affecting the outcome trend between the before- and after-intervention windows. 
Recent work has demonstrated that incorporating a continuous measure of exposure requires even stricter parallel trends assumptions \citep{Callaway2021Difference-in-DifferencesTreatment}.

In the DiD setting with a binary exposure variable, substantial methodology has been developed to relax the counterfactual parallel trends assumption by adjusting for observed confounders between the treated and control groups \citep{Abadie2005SemiparametricEstimators, Li2019, Santanna2020}.
In settings with a continuous exposure variable, the target parameter is no longer a single effect but a dose-response or causal effect \textit{curve}, which represents the average potential outcome or causal effect were all units given a particular exposure level.
However, few works have targeted confounding between different exposure levels, which occurs when assignment to different exposure levels is non-random by factors that also affect the outcome, when estimating effect curves in DiD studies.
For example, continuous exposures to an excise tax, such as distance to the city border or store-level pass-through rates, may be associated with population characteristics like socioeconomic status that are also associated with beverage sales. 
\cite{Han2019CausalChildren} considered DiD extensions of the approach developed by \cite{Hirano2004TheTreatments} to estimate causal effect curves for continuous exposures by utilizing an estimate for the generalized propensity score as a bias-adjusting covariate in a model regressing the outcome trends on exposure level.
However, such approaches hinge entirely on accurately specifying the conditional density function for the continuous exposure given covariates and often yield less efficient estimates than approaches that utilize outcome models even under correct density specification \citep{Austin2018AssessingOutcomes}.

Outside of DiD designs, there have been notable advancements in estimating dose-response curves for continuous exposures in the presence of confounding.
These include doubly robust estimators that utilize both generalized propensity scores and outcome regression models to estimate effect curves and generally improve upon approaches solely reliant on the generalized propensity score in terms of both efficiency and robustness \citep{Diaz2013TargetedCurve, Kennedy2017NonparametricEffects, Bonvini2022FastEstimation}.
Still, methods to harness these properties have not yet been developed to account for the multiple sources of confounding -- between treated and control groups \textit{and} between different exposure levels among the treated group -- and longitudinal design common in policy evaluations that employ a control group.

In this work, we develop novel estimators for causal effect curves that account for multiple sources of confounding under a DiD framework. 
Our methodology incorporates both outcome and treatment models to rigorously adjust for confounding effects and exploit the efficiency of semi-parametric influence-function based estimators.
Our estimators are multiply robust in the sense that only a subset of the four required nuisance function models need to be well-specified for consistent effect estimation.
To further mitigate potential model misspecification, our approach allows for flexible modeling techniques to capture complex confounding relationships and non-parametric effect curves.

The remainder of the paper is organized as follows: in Section~\ref{s:methods}, we introduce the setting of interest, identification assumptions, and proposed estimators. 
In Section~\ref{s:sims}, we demonstrate finite sample properties of our estimators under different nuisance function model specifications through simulation studies.
We then apply our methodology to study the effects of the Philadelphia beverage tax in Section~\ref{s:app} and conclude with a discussion in Section~\ref{s:discuss}. 

\section{Methods}
\label{s:methods}

\subsection{Setting}

We specifically consider the setting where a policy \textit{intervention} induces a distribution of \textit{exposures} on the population receiving the intervention.
We assume the intervention is introduced between two time-periods, $t=0,1$. 
Units, denoted with subscript $i$, are either in the intervention group, $A_i=1$, or not, $A_i=0$, and the time-varying intervention status at time $t$ is denoted by $Z_{it}=tA_i \in \{0,1\}$. 
Analogously, the inherent exposure ``assigned" to a unit is measured by $D_{i} \in \mathbbm{D}$, with the time-varying exposure status given by $\Omega_{it} = tD_i$.
Moving forward, we generally introduce formulation as if $\mathbbm{D} = \mathbbm{R}$ but our methodology also applies when $\mathbbm{D}$ is discrete.
Depending on our choice of exposure, a ``zero" exposure may not represent true protection from the intervention and therefore we denote this by $D_{i}=\emptyset$ or $\Omega_{it}=\emptyset$. 
We denote population-level vectors for intervention and exposure variables as $\mathbf{A}$, $\mathbf{Z_t}$, $\mathbf{D}$, and $\boldsymbol{\Omega_t}$, respectively. 
We observe outcomes of interest at each time point, $Y_{it}$. 
Finally, units and their outcomes may differ according to a vector of observed pre-intervention covariates, $\mathbf{X_i}$, although we could also include time-varying covariates assuming time-invariant effects \citep{Zeldow2021ConfoundingStudies}.
Then, we denote our observed data as $\mathbf{O_i} = (\mathbf{X_i}, A_i, D_i, Y_{i0}, Y_{i1})$.

The following assumptions allow us to define potential outcomes in terms of intervention and exposure status as $Y_{it}^{(Z_{it}, \Omega_{it})}$:
\begin{assumption}[Arrow of Time]\label{assump:arrow}
    $Y_{i0}^{(\mathbf{Z_0}, \boldsymbol{\Omega_0}, \mathbf{Z_1}, \boldsymbol{\Omega_1})} = Y_{i0}^{(\mathbf{Z_0}, \boldsymbol{\Omega_0})}$
\end{assumption}
\begin{assumption}[Stable Unit Treatment Value Assumption (SUTVA)]\label{assump:sutva}
$Y_{it}^{(\mathbf{Z_t}, \boldsymbol{\Omega_t})} = Y_{it}^{(Z_{it}, \Omega_{it})}$
\end{assumption}
Assumption~\ref{assump:arrow} states that potential outcomes do not depend on future treatment or exposure, which would be violated if, for example, units adapted their behavior in anticipation of an upcoming effect.
After connecting potential outcomes to treatment at a given time, we further assume that potential outcomes of a given unit only depend on population intervention and exposure status through a unit's own intervention and exposure status (Assumption~\ref{assump:sutva}). 
This formulation can be considered a relaxation of the traditional SUTVA for settings without a continuous exposure as we allow potential outcomes to vary with $\Omega_{it}$ in addition to $Z_{it}$.

We are primarily interested in estimating a curve for the causal effect of the intervention at different exposure levels, $\delta$, which we term the \textbf{A}verage \textbf{D}ose Effect on the \textbf{T}reated:
\begin{equation*}
    ADT(\delta) := \Psi(\delta) = E[Y_1^{(1,\delta)} - Y_1^{(0,\emptyset)} | A=1]
\end{equation*}
Here, the average is taken over each unit, so index $i$ is ommitted. This estimand addresses the question: What would be the average effect of a policy on the treated group where all treated units received the exposure of $\delta$? 

When estimating the Average Treatment Effect on the Treated (ATT), the most common target estimand in DiD studies with a binary intervention, we have a potential outcome, $Y_1^{(0,\emptyset)}$, unobserved for all treated units.
On the other hand, standard dose-response curves require identifying potential outcomes at a particular exposure level $\delta$, $Y_1^{(1,\delta)}$, which is unobserved for treated units $i$ where $D_i \neq \delta$.
To estimate the $ADT(\delta)$ in our setting, we must identify both sets of unobserved and partially observable potential outcomes.

\subsection{Identification}

To identify the $ADT(\delta)$, we require additional assumptions to link the unobservable and partially observable potential outcomes to observable data. 
First, we extend two standard causal inference assumptions to the continuous exposure setting:
\begin{assumption}[Consistency]\label{assump:consistency}
    $Y_{it}^{(z, \omega)} = Y_{it}$ when $(Z_{it}, \Omega_{it})=(z, \omega)$
\end{assumption}
\begin{assumption}[Positivity]\label{assump:pos} There exists $\epsilon > 0$ such that 
\begin{align*}
    &\text{(i) }\epsilon \leq \pi_A(\mathbf{x}) < 1 \text{ for all } \mathbf{x} \in \mathbbm{X} & [\pi_A(\mathbf{x}) = P(A=1|\mathbf{x})] \\ 
    &\text{(ii) } \epsilon \leq \pi_D(\delta|\mathbf{x}, A=1) \text{ for all } \mathbf{x} \in \mathbbm{X} | A=1\text{, } \delta \in \mathbbm{D}  & [\pi_D(\delta| A=1, \mathbf{x}) = p(D=\delta|A=1,\mathbf{X})]
\end{align*}
\end{assumption}
To estimate an effect curve, we require a stricter positivity assumption (Assumption~\ref{assump:pos}) than in the binary exposure setting, which only requires Assumption~\ref{assump:pos}(i).
Assumption~\ref{assump:pos}(ii) relates to covariate overlap between the different exposure levels and would be violated if certain subpopulations, defined by confounders $\mathbf{X}$, were restricted from certain exposure levels. 
Without appropriate covariate overlap, it is not possible to separate effect heterogeneity induced by the exposure from that induced by confounding.

A strong assumption underlying binary DiD approaches is that of counterfactual parallel trends, which allows us to connect $E[Y_1^{(0,\emptyset)}|A=1]$ to observable potential outcomes in the control group with $A=0$. 
While the DiD design innately adjusts for baseline differences between the treated and control groups by within-group differencing, it does not address situations when covariate distributions differ between treated and control groups and those covariates also affect outcome trends over time.
Here, we have adopted the approach others have taken to relax this assumption by assuming parallel trends holds after conditioning on a set of observed baseline confounders \citep{Abadie2005SemiparametricEstimators, Santanna2020}:
\begin{assumption}[Conditional Counterfactual Parallel Trends Between Treated and Control]\label{assump:ptA}
\centering $E[Y_{1}^{(0,\emptyset)}-Y_{0}^{(0,\emptyset)} | A=1,\mathbf{X}] = E[Y_{1}^{(0,\emptyset)}-Y_{0}^{(0,\emptyset)} | A=0,\mathbf{X}]$
\end{assumption}

\begin{figure}
\centerline{\includegraphics[width=6.25in]{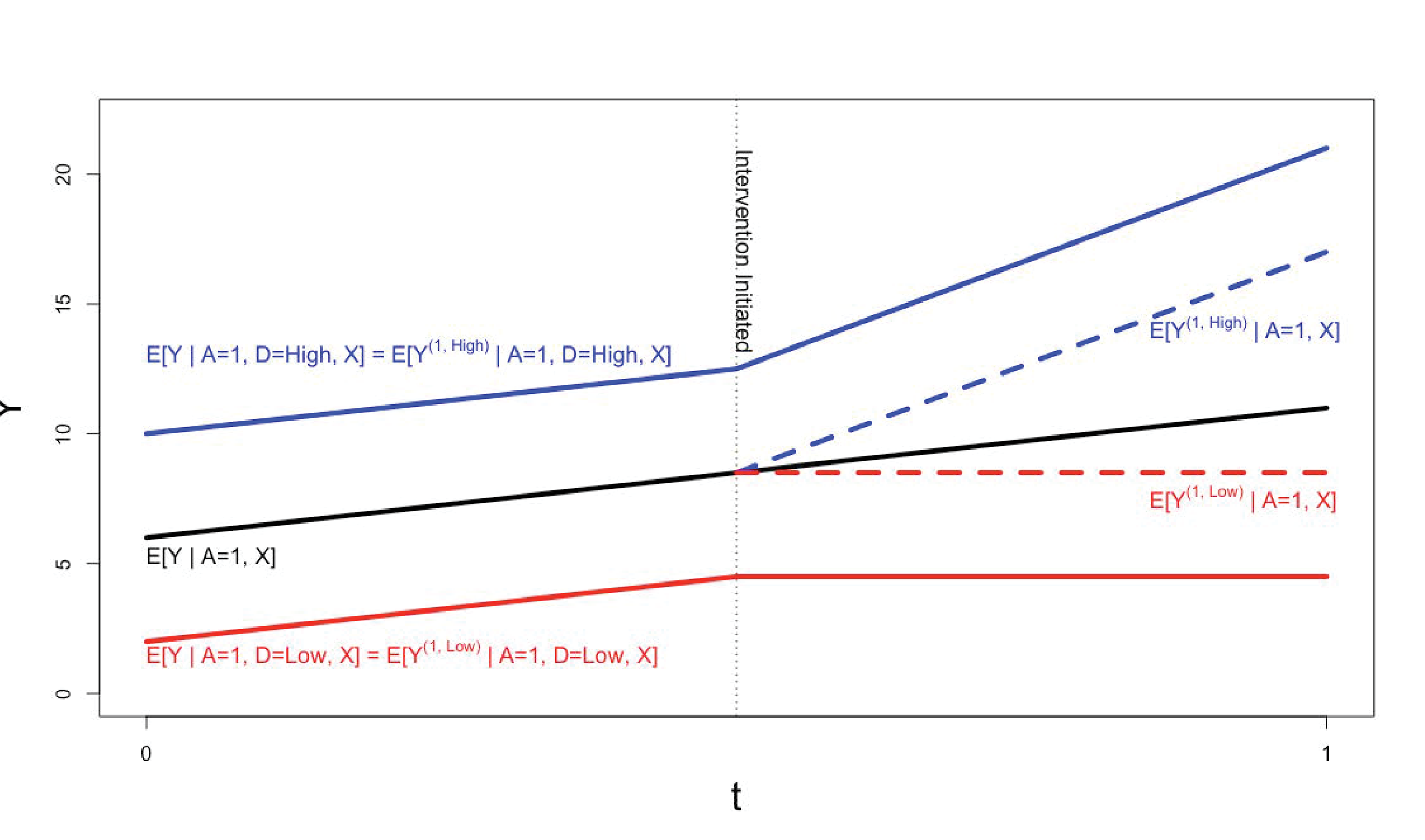}}
\caption{An example of Assumption~\ref{assump:ptD} where exposure level is discretized into $\mathbbm{D}=\{high,low\}$ for visualization.}
\label{fig:ptD}
\end{figure}

To connect $E[Y_1^{(1,\delta)}|A=1]$ to observable potential outcomes for all treated units, we require an additional parallel trends assumption:
\begin{assumption}[Conditional Counterfactual Parallel Trends Among Treated Between Doses]\label{assump:ptD}
\centering
$E[Y_{1}^{(1,\delta)}-Y_{0}^{(0,\emptyset)} | A=1,D=\delta,\mathbf{X}] = E[Y_{1}^{(1,\delta)}-Y_{0}^{(0,\emptyset)} | A=1,\mathbf{X}] \text{ for all } \delta \in \mathbbm{D}$
\end{assumption}
For intuition, consider similar units, as defined by $\mathbf{X}$, between (i) the subset of the treated group that receives a particular exposure level, $\delta$, and (ii) the entire treated group regardless of exposure level.
In principle, we require that the observed trends in group (i) are representative of the counterfactual trends in group (ii) were all of group (ii) intervened on with exposure level $\delta$. 
A visual example of this assumption is presented in Figure~\ref{fig:ptD}.
Previous works in the continuous exposure DiD setting have made similar but likely stronger assumptions, requiring this version of parallel trends holds unconditionally or that the trends in potential outcomes are independent of exposure given $\mathbf{X}$~\citep{Callaway2021Difference-in-DifferencesTreatment, Han2019CausalChildren}.

Assuming these conditions, we can identify the $ADT(\delta)$ with observable data:
\begin{theorem}[Identifiability]\label{thm:ident}
Under Assumptions~\ref{assump:arrow}-\ref{assump:ptD}, the $ADT(\delta)$ is identifiable through $\Psi(\delta) = E[\mu_{1,\Delta}(\delta,\mathbf{X}) - \mu_{0,\Delta}(\mathbf{X}) | A=1]$, where $\mu_{1,\Delta}(\delta,\mathbf{X}) = E[Y_1 - Y_0 | A=1, D=\delta, \mathbf{X}]$ and $\mu_{0,\Delta}(\mathbf{X}) = E[Y_1 - Y_0 | A=0, \mathbf{X}]$. 
\end{theorem}
A proof is provided in Web Appendix A. While assumptions regarding parallel trends are by definition untestable using observed data, it is common for practitioners to examine counterfactual parallel trends between the treated and control groups (Assumption~\ref{assump:ptA}) leading up to the intervention when multiple pre-treatment observations are available (e.g., $t=-1$) under a binary treatment \citep{Bilinski2018}.
The test derives its usefulness by assuming the fully observable relationship between $E[Y_{0}^{(0,\emptyset)} - Y_{-1}^{(0,\emptyset)}|A=1]$ and $E[Y_{0}^{(0,\emptyset)} - Y_{-1}^{(0,\emptyset)}|A=0]$ continues and is therefore a reasonable proxy for the unobservable relationship between $E[Y_{1}^{(0,\emptyset)} - Y_{0}^{(0,\emptyset)}|A=1]$ and $E[Y_{1}^{(0,\emptyset)} - Y_{0}^{(0,\emptyset)}|A=0]$. 
Similar placebo tests are possible for Assumption~\ref{assump:ptD}; however, $E[Y_t^{(1,\delta)}|A=1]$ is unobservable in the pre-treatment period since intervention and exposure are inactive. 
Therefore, the usefulness of comparable tests assessing $E[Y_{0}^{(0,\emptyset)} - Y_{-1}^{(0,\emptyset)}|A=1, D=\delta] = E[Y_{0}^{(0,\emptyset)} - Y_{-1}^{(0,\emptyset)}|A=1]$ also depend on how well these trends represent counterfactual trends at different exposure levels.

\subsection{Multiply-Robust Estimator}

Our goal is to find an estimator for the effect curve, $ADT(\delta)$. 
A potentially ideal estimator for the $ADT(\delta)$ would be based on the efficient influence function (EIF) for this estimand, as EIF-based estimators come with optimal convergence properties and often, though not necessarily, certain robustness to model misspecification. 
However, the EIF is not tractable for the $ADT(\delta)$ without requiring restrictive parametric assumptions on the form of the effect curve as the infinite dimensionality of a non-parametric curve does not satisfy the requirement for pathwise differentiability \citep{Kennedy2022SemiparametricReview}.
Therefore, we first consider an aggregate functional, $\Psi$, which integrates the $ADT(\delta)$ over the marginal density of exposures, $f$:
\begin{equation*}
\Psi = \int\limits_\mathbbm{D} E[Y_1^{(1,\delta)} - Y_1^{(0,\emptyset)} | A=1] df(\delta|A=1) = \int\limits_\mathbbm{D} ADT(\delta) df(\delta|A=1)
\end{equation*}
Importantly, this functional, as an integration of the $ADT(\delta)$, is pathwise differentiable and therefore has a tractable EIF~\citep{Kennedy2022SemiparametricReview}.
\begin{theorem}[EIF for $\Psi$]\label{thm:if}
The efficient influence function for $\Psi$ is given by:
$$\phi(\mathbf{X}, A, D, Y_0, Y_1) = \frac{A}{P(A=1)} \xi(\mathbf{X}, A, D, Y_0, Y_1) +  \tau(\mathbf{X}, A, Y_0, Y_1) + J(\mathbf{X}, A) - \Psi$$

Where 
\begin{align*}
    &\xi(\mathbf{X}, A, D, Y_0, Y_1; \mu_{1,\Delta}, \pi_D) = m(D|A=1) + \frac{(Y_1-Y_0) - \mu_{1,\Delta}(D,\mathbf{X})}{\pi_D(D|A=1,\mathbf{X})} f(D|A=1) \\
    &\tau(\mathbf{X}, A, Y_0, Y_1; \mu_{0,\Delta}, \pi_A) = \frac{A}{P(A=1)}\mu_{0,\Delta}(\mathbf{X}) + \frac{(1-A)\pi_A(\mathbf{X})[(Y_1 - Y_0) - \mu_{0,\Delta}(\mathbf{X})]}{P(A=1)(1-\pi_A(\mathbf{X}))} \\
    &J(\mathbf{X}, A; \mu_{1,\Delta}, \pi_D) = \frac{A}{P(A=1)} \int\limits_\mathbbm{D} \{ \mu_{1,\Delta}(d, \mathbf{X}) - m(d|A=1) \} df(d|A=1)\\
    &m(D|A=1) = \int\limits_\mathbbm{X} \mu_{1,\Delta}(D,\mathbf{x}) dP(\mathbf{x}|A=1)\\
    &f(D|A=1) = \int\limits_\mathbbm{X} \pi_D(D|A=1,\mathbf{x}) dP(\mathbf{x}|A=1)
\end{align*}
\end{theorem}

A proof is provided in Web Appendix B. 
While notationally complex, the above influence function can be decomposed into three parts, each of which has an interpretable role. 
First, note that $\xi(\mathbf{X}, A, D, Y_0, Y_1)$  corresponds to the dose-specific component, $E[Y_1-Y_0|A=1,D=D]=E[Y_1(1,D)-Y_0(0,\emptyset)|A=1]=:\theta(D)$.
Here, $m(D|A=1)$ averages the conditional expectation, $\mu_{1,\Delta}(D, \mathbf{X})=E[Y_1-Y_0|A=1,D,\mathbf{X}]$, across all covariates in the treated group, giving us an outcome model-based formulation of $\theta(D)$. 
The second term of $\xi$ re-weights observed deviations between observed and expected outcome trends by the balance-inducing generalized propensity score weight, $w^{(1)} = f(D|A=1)/\pi_D(D|A=1,\mathbf{X})$, as a sort of bias correction term when $\mu_{1,\Delta}$ is misspecified.
Similarly, $\tau(\mathbf{X}, A, Y_0, Y_1)$ corresponds to the control-specific component, $E[Y_1(0,\emptyset)-Y_0(0,\emptyset)|A=1]=:\theta_0$.
The first term of $\tau$ provides estimates for $\mu_{0, \Delta}(\mathbf{X})=E[Y_1-Y_0|A=0,\mathbf{X}]$ for all treated units.
The second term of $\tau$ re-weights observed deviations between observed and expected outcome trends for the control group by ATT propensity score weights, $w^{(0)}=\pi_A(\mathbf{X})/(1-\pi_A(\mathbf{X}))$, serving as a bias correction term in case $\mu_{0,\Delta}$ is misspecified. 
Finally, we note that $J(\mathbf{X}, A)$ is independent of exposure dose and $E[J(\mathbf{X}, A)|A=1]=0$, so will not factor into point estimation for either $\Psi(\delta)$ or $\Psi$.
A similar approach was used by \cite{Kennedy2017NonparametricEffects} to decompose $\theta(D)$ and generate their so-called pseudo-outcomes when estimating a dose-response curve in cross-sectional studies.

To utilize this influence function and subsequent decomposition for estimation of $\Psi(\delta)$, we require a multi-step approach as follows:
\begin{enumerate}[1.]
\item Estimate nuisance functions $\hat{\mu}_{1,\Delta}, \hat{\pi}_D, \hat{\mu}_{0,\Delta}, \hat{\pi}_A$.
\item Plug in data and estimated nuisance functions to construct $\hat{\xi}(\mathbf{O};\hat{\pi}_D,\hat{\mu}_{1,\Delta})$ and $\hat{\tau}(\mathbf{O};\hat{\pi}_A,\hat{\mu}_{0,\Delta})$.
\item Regress $\hat{\xi}$ on dose variable $D$ to obtain $\hat{\theta}(\delta)$ and take the empirical mean of $\hat{\tau}$ as $\hat{\theta}_0$. 
\item Set $\hat{\Psi}(\delta) = \hat{\theta}(\delta) - \hat{\theta}_0$.
\end{enumerate}
The key difference between estimation of $\Psi(\delta)$ and a traditional EIF-based estimator is the required estimation of $E[\xi | A=1, D]$, rather than $E[\xi | A=1]$, in Step (3). 
To estimate the former, we advocate for a flexible non-parametric regression technique, like the local linear kernel regression, to minimize bias induced by parametric assumptions \citep{Wasserman2006AllStatistics}.

Practically speaking, $\hat{\pi}_D$ and $\hat{\mu}_{1,\Delta}$ must be estimated with data from the $A=1$ group whereas $\hat{\pi}_A$ requires data from both $A=1$ and $A=0$ groups. 
Estimation of $\hat{\mu}_{0,\Delta}$ could incorporate data from both $A=1$ and $A=0$ groups, but the efficiency gained from utilizing $A=1$ data comes at the cost of robustness in that correctly specifying $\hat{\mu}_{0,\Delta}$ would require correctly modeling treatment effect dynamics, linking its specification to that of $\hat{\mu}_{1,\Delta}$. 
Another helpful practical step is to normalize $w^{(1)} | A=1$ and $w^{(0)} | A=0$ so that  both have mean one (e.g., Hajek weighted estimators), which has been shown to improve the stability of doubly robust DiD estimators for binary exposures \citep{Santanna2020}.

Our decomposition of the influence function $\phi$ also foreshadowed important robustness properties, which can be formalized in the following theorem:
\begin{theorem}[Multiple Robustness]\label{thm:robust}
Suppose that Assumptions~\ref{assump:arrow}-\ref{assump:ptD} hold and:
\begin{itemize}
    \item[(a)] $\hat{\pi}_D \xrightarrow[]{p} \pi_D$ OR $\hat{\mu}_{1,\Delta} \xrightarrow[]{p} \mu_{1,\Delta}$.
    \item[(b)] $\hat{\pi}_A \xrightarrow[]{p} \pi_A$ OR $\hat{\mu}_{0,\Delta} \xrightarrow[]{p} \mu_{0,\Delta}$.
    \item[(c)]  $(\hat{\pi}_D,\hat{\mu}_{1,\Delta})$ and their limits are constrained in uniformly bounded function classes with finite uniform entropy integrals, with $1/\hat{\pi}_D$ and its limit as $n_A \rightarrow \infty$ also uniformly bounded.
    \item[(d)]  The bandwidth of the kernel regression, $h=h_n$, satisfies $h \rightarrow 0$ and $n_Ah^3 \rightarrow \infty$.
    \item[(e)]  The kernel function, $K$, is a continuous symmetric probability density with support $[-1,1]$.
    \item[(f)]  $\Psi(\delta)$ is twice continuously differentiable and both $f(\delta | A=1)$ and the conditional density of $\hat{\xi}$ as $n_A \rightarrow \infty$ given $D=\delta,A=1$ are continuous as functions of $\delta$.
\end{itemize}
Then, $\hat{\Psi}(\delta) \xrightarrow{p} \Psi(\delta)$ for an exposure value, $\delta$, in the interior of the compact support of $D$ as $n\rightarrow\infty$ and $n_A=\sum\limits_i A_i \rightarrow \infty$.
\end{theorem}
A proof is provided in Web Appendix C. 
Condition (c) puts a mild restriction on the flexibility of the dose-specific nuisance estimators and their corresponding limits, but still allows for the use of several non-parametric estimators and can be relaxed further using sample-splitting techniques~\citep{vanderLaan2011TargetedLearning}. 
Conditions (d)-(f) are specific to the local linear kernel regression estimator, but are considered quite weak \citep{Kennedy2017NonparametricEffects}.
Such flexibility comes at the cost of convergence rates, which are provided in Web Appendix D and demonstrated in Section~\ref{s:sims}.

Importantly, we only require one correct model specification in each specified pair of nuisance functions corresponding to $\theta(\delta)$ and $\theta_0$, respectively.
Accordingly, our estimator will result in consistent effect estimation in 9 of the 16 ($56\%$) permutations of correct/incorrect nuisance function specifications (i.e., \{$\pi_D \in \{\text{Correct,Incorrect}\}, \mu_{1,\Delta}\in \{\text{Correct,Incorrect}\}, \pi_A\in \{\text{Correct,Incorrect}\}, \mu_{0,\Delta}\in \{\text{Correct,Incorrect}\}$\}).
This rate is lower than the $75\%$ (3 of 4 permutations) consistency rate from standard doubly robust estimators which only require two nuisance functions, a necessary cost to address multiple levels of confounding in terms of robustness.
Still, if we consider the specification of both outcome models, $\{\mu_{1,\Delta}, \mu_{0,\Delta}\}$, together and both propensity score models, $\{\pi_D, \pi_A\}$, together, then we have consistent estimates in 3 of the 4 specification permutations and further achieve consistency under additional ``partial” misspecifications.

\subsection{Uncertainty Quantification}

Closed-form inference for $\hat{\Psi}(\delta)$ is not immediately derivable for several reasons. First, $\hat{\Psi}(\delta)$ does not absorb the semi-parametric efficiency guarantees of influence-function based estimators that doubly robust estimators of binary treatments achieve under correct nuisance function specification.
Second, $\hat{\Psi}(\delta)$ is asymptotically normal under mild restrictions on nuisance function convergence rates (Web Appendix D), but these asymptotic assumptions ignore finite sample uncertainty and thus likely  substantially underestimate the true variance in practical settings \citep{Lunceford2004StratificationStudy}.
Third, while $\hat{\Psi}(\delta)$ is a consistent estimator of $\Psi(\delta)$, it is centered about a smoothed version of the true effect curve, leading to potential under-coverage of desired confidence intervals \citep{Wasserman2006AllStatistics}.
This last challenge is standard in non-parametric regression and requires either under-smoothing of the bandwidth, which is more of a technical device than a practical tool, or computationally-intensive techniques that appear impractical in our multi-step estimation setting where nuisance function models must also be estimated \citep{Racine2004NonparametricData}.
Therefore, we follow others in expressing pointwise uncertainty about this smoothed curve \citep{Kennedy2017NonparametricEffects}.
However, we can improve upon the rigid assumptions and finite sample performance of standard asymptotic approximations by deriving sandwich variance estimators or utilizing bootstrapping approaches for pointwise confidence bands.

To derive sandwich variance estimators, we represent our plug-in estimator as a series of unbiased estimating equations and apply M-estimation theory to engineer sandwich estimators for the variance of estimated parameters \citep{Stefanski2002TheM-Estimation}.
Concretely, we first define the set of estimating equations, $\sum\limits_i \boldsymbol{\Gamma_i}(\boldsymbol{\eta}, \delta)=0$, where $\boldsymbol{\Gamma_i}(\boldsymbol{\eta}, \delta)=$
\begin{equation*}\left[\begin{aligned}
    & \frac{A_i}{P(A=1)} \{ K(\frac{D_i-\delta}{h})(\hat{\xi}_i - \hat{\theta}(\delta) - (\frac{D_i-\delta}{h}) \hat{\beta}) \\
    & \qquad + \int\limits_\mathbbm{D} K(\frac{D_i-\delta}{h}) (\hat{\mu}_{1,\Delta}(\mathbf{X_i},d) - \hat{m}(d|A=1)) d\hat{f}(d|A=1)\} \\
    & \frac{A_i}{P(A=1)} \{ (\frac{D_i-\delta}{h}) K(\frac{D_i-\delta}{h})(\hat{\xi}_i - \hat{\theta}(\delta) - (\frac{D_i-\delta}{h}) \hat{\beta}) \\
    & \qquad + \int\limits_\mathbbm{D} (\frac{D_i-\delta}{h}) K(\frac{D_i-\delta}{h}) (\hat{\mu}_{1,\Delta}(\mathbf{X_i},d) - \hat{m}(d|A=1)) d\hat{f}(d|A=1)\} \\
    & (1-A_i) \{ \hat{\theta}_{00} - \frac{1}{P(A=1)} \frac{\hat{\pi}_A(\mathbf{X_i})}{1-\hat{\pi}_A(\mathbf{X_i})}(Y_{i1}-Y_{i0}-\hat{\mu}_{0,\Delta}(\mathbf{X_i})) \} \\
    & A_i \{ \hat{\theta}_{01} - \frac{1}{P(A=1)} \hat{\mu}_{0,\Delta}(\mathbf{X_i}) \}
  \end{aligned}\right]
  \end{equation*}
where $\hat{\boldsymbol{\eta}} = (\hat{\theta}(\delta), \hat{\beta}, \hat{\theta}_{00}=E_n[\frac{1-A}{P(A=1)} \frac{\hat{\pi}_A(\mathbf{X})}{1-\hat{\pi}_A(\mathbf{X})}(Y_1-Y_0-\hat{\mu}_{0,\Delta}(\mathbf{X}))], \hat{\theta}_{01}=E_n[\frac{A}{P(A=1)}\hat{\mu}_{0,\Delta}(\mathbf{X})])^T$ are estimable parameters based on the efficient influence functions for local linear kernel regression (first two entries) and $\Theta_0$ (last two entries) \citep{Kennedy2017NonparametricEffects}.
Under the existence of $E[\frac{\partial}{\partial \boldsymbol{\eta}^{\prime}} \boldsymbol{\Gamma}(\boldsymbol{\eta})]$, $E[\boldsymbol{\Gamma}(\boldsymbol{\eta})\boldsymbol{\Gamma}(\boldsymbol{\eta})^T]$, and their sample estimators with $\hat{\boldsymbol{\eta}}$, we can then use the sandwich variance formula to estimate the covariance matrix for $\boldsymbol{\eta}$ as $\widehat{V} = \{ \sum\limits_{i=1}^n \frac{\partial}{\partial \boldsymbol{\eta}^{\prime}} \boldsymbol{\Gamma_i}(\hat{\boldsymbol{\eta}}) \}^{-1} \{ \sum\limits_{i=1}^n \boldsymbol{\Gamma_i}^{\bigotimes 2}(\hat{\boldsymbol{\eta}}) \} \{ \sum\limits_{i=1}^n \frac{\partial}{\partial \boldsymbol{\eta}} \boldsymbol{\Gamma_i}^{\prime}(\hat{\boldsymbol{\eta}}) \}^{-1}$.
Finally, we can calculate $\widehat{Var}(\hat{\Psi}(\delta)) = \widehat{Var}(\hat{\theta}(\delta) - \hat{\theta}_{00} - \hat{\theta}_{01}) = \hat{V}_{1,1} + \hat{V}_{3,3} + \hat{V}_{4,4} + 2(\hat{V}_{3,4} - \hat{V}_{1,3} - \hat{V}_{1,4})$ and derive $95\%$ confidence intervals for each $\delta$ under a normal approximation.

Still, this sandwich estimator relies on a normal approximation and is not readily extensible to alternative parameters of interest (e.g., $\hat{\Psi}(\delta_1) - \hat{\Psi}(\delta_2)$). 
Further, while we demonstrate potential extensions to the proposed sandwich variance estimator to incorporate uncertainty in nuisance function estimation in Web Appendix E, such extensions must be derived for a specific set of nuisance function models and may not be possible for all choices, inhibiting method flexibility. 
Alternatively, nonparametric bootstrapping approaches can flexibly be used for many parameters of interest while incorporating nuisance function estimation \citep{Wasserman2006AllStatistics, Funk2011DoublyEffects}.
Here, we consider a non-parametric weighted bootstrap where we sample unit-specific, rather than observation-specific, weights to account for correlation between multiple observations of the same unit \citep{Li2019}.
After sampling these weights from an $exponential(1)$ distribution, we then scale them such that the sum of weights in each intervention group is equal to their observed sample sizes and utilize these weights throughout the estimation process. 
Allowing for small but non-zero weights, unlike a discrete bootstrap sampling approach, is particularly helpful for improving the observed support of $D$ within bootstrap samples.

\section{Simulation Studies}
\label{s:sims}

\subsection{Data Generation and Methods}

\begin{figure}
\centerline{\includegraphics[width=6.25in]{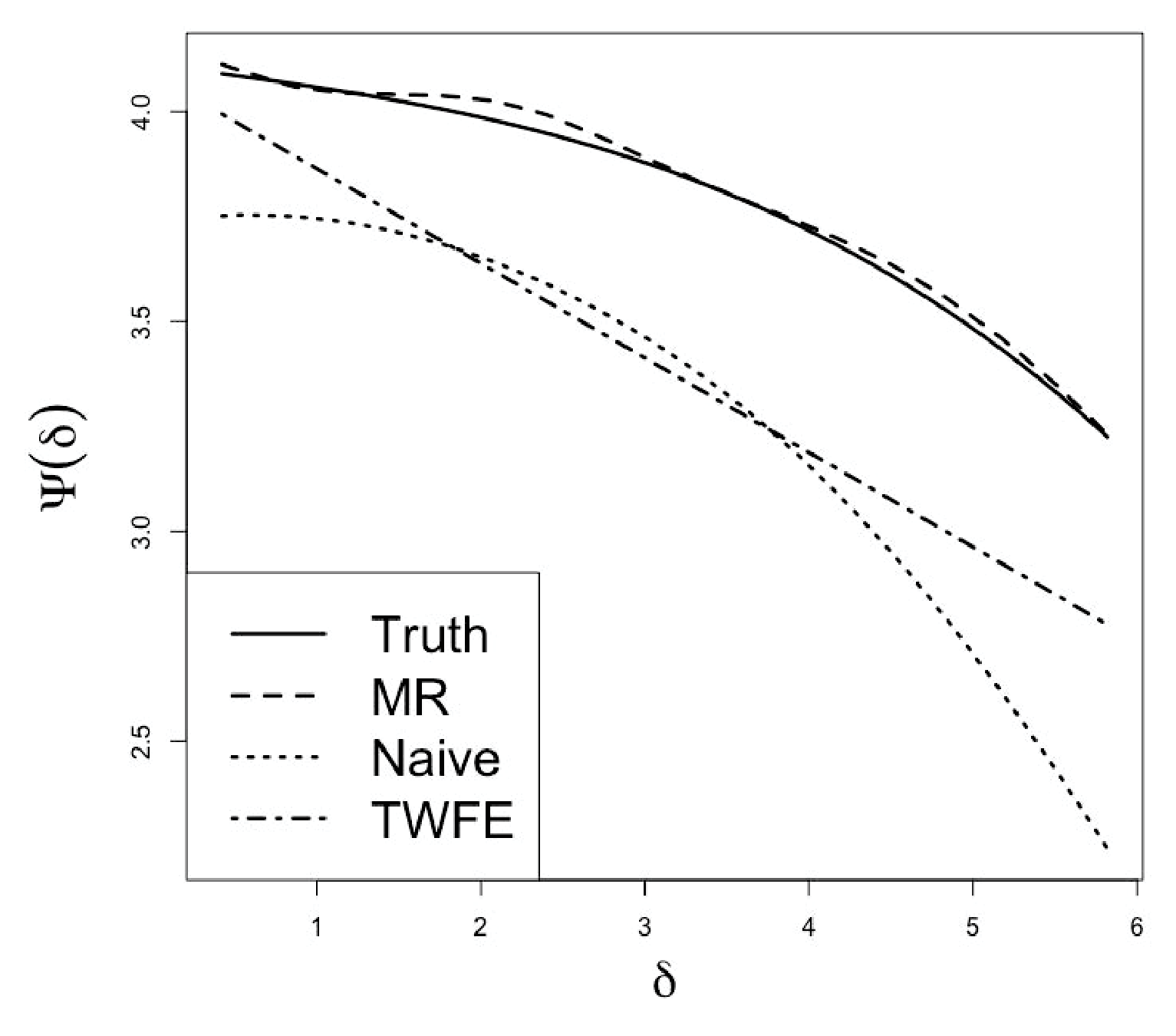}}
\caption{An example fit for one simulated dataset ($n=1000$) showing the true effect curve and estimated effect curves using the multiply robust (MR), confounding-naive (Naive), and two-way fixed effects (TWFE) approaches. A constant difference between the MR and Naive approaches represents confounding between treated and control groups while the increasing difference over values of $\delta$ represents confounding between units at different exposure levels.}
\label{fig:sims}
\end{figure}

We conducted numerical studies to examine the finite-sample properties of our proposed methods relative to alternative approaches under different forms of model misspecification. Specifically, to generally mirror the structure of the data from our motivating study presented in Section~\ref{s:app}, we simulated normally distributed covariates,
\begin{equation*}
    \mathbf{X_i}=(X_{i1},X_{i2},X_{i3}, X_{i4})^T \sim N(\mathbf{0}, \mathbf{I}_4)
\end{equation*}
a binary intervention,
\begin{align*}
    &A_i\mid \mathbf{X_i} \sim Bern(expit(-0.1 + 0.05X_{i1}+0.05X_{i2}-0.05X_{i3}+0.15X_{i4}))
\end{align*}
where $expit(z)=exp(z)/[1+exp(z)]$, a normally distributed exposure among treated units,
\begin{equation*}
    D_i\mid \mathbf{X_i}, A_{i}=1 \sim N(\mu=3+0.2X_{i1}+0.25X_{i2}-0.3X_{i3}+0.5X_{i4}, \sigma^2=4)
\end{equation*}
a normally distributed outcome at time $t=0$,
\begin{equation*}
    Y_{i0} \mid \mathbf{X_i}, A_i \sim N(\mu=10+0.4X_{i1}-X_{i2}+0.4X_{i3}+0.3X_{i4}+2A_i, \sigma^2=0.3^2)
\end{equation*}
and a normally distributed outcome at time $t=1$ where the expected trend is determined by confounders, intervention group, and exposure level,
\begin{align*}
    &Y_{i1} \mid \mathbf{X_i}, A_i, D_i \sim N(\mu=Y_{i0}+A_i\lambda_1(\mathbf{X_i}, D_i) + (1-A_i)\lambda_0(\mathbf{X_i}), \sigma^2=0.7^2) \\
    &\lambda_1(\mathbf{X_i}, D_i) = 1 + 0.6X_{i1}+0.6X_{i2}+0.9X_{i3}-0.3X_{i4}+2A_i \\
    & \qquad + D(0.04 - 0.1X_{i1}+0.1X_{i3} - 0.03/10D^2)\\
    &\lambda_0(\mathbf{X_i}) = -3 -X_{i1}+0.7X_{i2}+0.6X_{i3}-0.6X_{i4}
\end{align*}
The ground truth effect curve, as shown in the solid line in Figure~\ref{fig:sims}, was then calculated at 50 evenly-spaced exposure values between the 10th and 90th percentiles using Monte Carlo integration over a super-population of size $n=1,000,000$ as $\Psi(\delta) = \int\limits_\mathbbm{X} [\lambda_1(\mathbf{x}, \delta) - \lambda_0(\mathbf{x}, \delta)] dP(\mathbf{x}|A=1)$.

Using the simulated data, we compared our proposed, multiply robust method (MR) to four alternative estimators that one would consider applying in practice and demonstrate the different aspects of robustness.
First, we estimated the standard linear two-way fixed effects (TWFE) model, $Y_{it} = \beta_0 + \boldsymbol{\beta}^{T}\mathbf{X_i} + \gamma_T t + \gamma_A A_i + \gamma_D A_i D_i + \tau_0 t A_i + \tau_D t A_i D_i + \epsilon_{it}$ studied by \cite{Callaway2021Difference-in-DifferencesTreatment} that does not adjust for confounding between intervention or exposure levels and outcome trends, and calculated an effect curve as $\hat{\Psi}^{(twfe)}(\delta)=\hat{\tau}_0 + \hat{\tau}_D \delta$. 
Second, we estimated a flexible, confounding-naive difference-in-differences estimator as $\hat{\Psi}^{(naive)}(\delta)=\hat{E}[Y_1-Y_0|D=\delta,A=1] - \hat{E}[Y_1-Y_0|A=0]$, where the first expectation was estimated with a local linear kernel regression where the bandwidth parameter is selected via cross-validation and the second with a sample mean.
Then, we considered outcome regression and IPW analogies to our proposed estimator.
The outcome regression approach relies entirely on correct specification of $\mu_{1,\Delta}$ and $\mu_{0,\Delta}$ and estimates $\hat{\Psi}^{(or)}(\delta) = \hat{E} [\hat{\mu}_{1,\Delta}(\mathbf{X}) - \hat{\mu}_{0,\Delta}(\mathbf{X}) | A=1]$ with a sample mean taken over treated units.
The IPW approach relies entirely on specification of $\pi_{D}$ and $\pi_{A}$, estimating $\hat{\theta}^{(ipw)}(\delta) = \hat{E}[\hat{w}^{(1)}(Y_1-Y_0)|A=1, D=\delta]$ with a local linear kernel regression and $\hat{\theta}^{(ipw)}_0 = \hat{E}[\hat{w}^{(0)}(Y_1-Y_0)|A=0]$ with a sample mean taken over control units, to get $\hat{\Psi}^{(ipw)}(\delta) = \hat{\theta}^{(ipw)}(\delta) - \hat{\theta}^{(ipw)}_0$.
This is a similar approach to that of \cite{Han2019CausalChildren} but using estimated propensity score-based weights rather than covariate adjustment.

We considered nuisance function misspecification by developing models under the specified mean functions, where we used $\mathbf{X}$ for covariates in our model under a correct specification and $\mathbf{W}$, the covariate transformation of $\mathbf{X}$ from \cite{KangSchafer2007}, for covariates under an incorrect specification. 
For each of the 16 permutations of nuisance model specification, we then generated 1000 simulated datasets for each of three sample sizes: $n=200$, $1000$, and $10000$.
Potential estimators were compared relative to the ground truth effect curve using integrated absolute mean bias and root mean squared error (RMSE), where integrations were taken across the 1000 simulations and then according to the ground truth marginal density of $D$ as calculated empirically from the super-population. 
Finally, we assessed our finite sample inferential approaches by calculating the integrated coverage probabilities and interval width of confidence intervals derived from the described sandwich and bootstrap variance approaches for our proposed multiply robust estimator, where we used the $2.5$ and $97.5$ percentiles among 500 bootstrap replicates as our interval bounds.

\subsection{Results}

Results summarizing the performance of causal effect curve point estimates are given in Table~\ref{t:sims_main} $(n=1000)$ and Tables A1-A2 $(n=200,$ $5000)$. 
In all analyzed scenarios, $\hat{\Psi}^{(twfe)}(\delta)$ and $\hat{\Psi}^{(naive)}(\delta)$ exhibit significant bias as they fail to account for confounding between the treated and control groups and between the different exposure levels within the treated group.
When $\hat{\pi}_A$ or $\hat{\pi}_D$ are misspecified ($75\%$ of scenarios), $\hat{\Psi}^{(ipw)}(\delta)$ demonstrates bias, the severity of which depends upon the specific models that are misspecified.
Similarly, $\hat{\Psi}^{(or)}(\delta)$ is biased in the $75\%$ of scenarios when $\hat{\mu}_{0,\Delta}$ or $\hat{\mu}_{1,\Delta}$ are misspecified. 
Notably, our proposed estimator, $\hat{\Psi}^{(mr)}(\delta)$, exhibits bias only when both $\hat{\pi}_A$ and $\hat{\mu}_{0,\Delta}$ \textit{or} both $\hat{\pi}_D$ and $\hat{\mu}_{1,\Delta}$ are misspecified ($44\%$ of scenarios). 
Further, $\hat{\Psi}^{(mr)}(\delta)$ correctly accounts for at least one of the two sources of confounding as long as one of the four nuisance functions are well-specified. 

\begin{table}
\caption{Comparison of Bias (RMSE) of effect curve estimates under multiply robust (MR), outcome regression (OR), and inverse probability weighting (IPW) approaches in simulation scenarios varying by nuisance function specification ($n=1000$). Green cells represent scenarios where both sources of confounding are correctly adjusted for asymptotically, whereas orange and red cells represent scenarios where one and both sources are not correctly adjusted for, respectively. Confounding-naive and two-way fixed effects estimators have Bias (RMSE) of 0.324 (0.136) and 0.418 (0.200) in all scenarios.}
\label{t:sims_main}
\begin{center}
\begin{tabular}{lccc}
\hline
Incorrect Models & MR & OR & IPW \\
\hline
None & \cellcolor{green!40} 0.027 (0.010) & \cellcolor{green!40} 0.001 (0.008) & \cellcolor{green!40} 0.028 (0.016) \\
$\pi_A$ & \cellcolor{green!40} 0.027 (0.010) & \cellcolor{green!40} 0.001 (0.008) & \cellcolor{orange!40} 0.129 (0.035) \\
$\mu_{0,\Delta}$ & \cellcolor{green!40} 0.013 (0.011) & \cellcolor{orange!40} 0.094 (0.021) & \cellcolor{green!40} 0.028 (0.016) \\
$\pi_D$ & \cellcolor{green!40} 0.027 (0.010) & \cellcolor{green!40} 0.001 (0.008) & \cellcolor{orange!40} 0.126 (0.040) \\
$\mu_{1,\Delta}$ & \cellcolor{green!40} 0.029 (0.016) & \cellcolor{orange!40} 0.181 (0.051) & \cellcolor{green!40} 0.028 (0.016) \\
$\pi_A, \pi_D$ & \cellcolor{green!40} 0.027 (0.010) & \cellcolor{green!40} 0.001 (0.008) & \cellcolor{red!40} 0.165 (0.063) \\
$\mu_{0,\Delta}, \mu_{1,\Delta}$ & \cellcolor{green!40} 0.016 (0.017) & \cellcolor{red!40} 0.232 (0.082) & \cellcolor{green!40} 0.028 (0.016) \\
$\mu_{0,\Delta}, \pi_D$ & \cellcolor{green!40} 0.013 (0.011) & \cellcolor{orange!40} 0.094 (0.021) & \cellcolor{orange!40} 0.126 (0.040) \\
$\pi_A, \mu_{1,\Delta}$ & \cellcolor{green!40} 0.029 (0.016) & \cellcolor{orange!40} 0.181 (0.051) & \cellcolor{orange!40} 0.129 (0.035) \\
$\pi_A, \mu_{0,\Delta}$ & \cellcolor{green!40} 0.139 (0.032) & \cellcolor{orange!40} 0.094 (0.021) & \cellcolor{orange!40} 0.129 (0.035) \\
$\pi_D, \mu_{1, \Delta}$ & \cellcolor{orange!40} 0.128 (0.039) & \cellcolor{orange!40} 0.181 (0.051) & \cellcolor{orange!40} 0.126 (0.040) \\
$\pi_A, \mu_{0,\Delta}, \pi_D$ & \cellcolor{orange!40} 0.139 (0.032) & \cellcolor{orange!40} 0.094 (0.021) & \cellcolor{red!40} 0.165 (0.063) \\
$\pi_A, \mu_{0,\Delta}, \mu_{1,\Delta}$ & \cellcolor{orange!40} 0.139 (0.038) & \cellcolor{red!40} 0.232 (0.082) & \cellcolor{orange!40} 0.129 (0.035) \\
$\pi_A, \pi_D, \mu_{1, \Delta}$ & \cellcolor{orange!40} 0.128 (0.039) & \cellcolor{orange!40} 0.181 (0.051) & \cellcolor{red!40} 0.165 (0.063) \\
$\mu_{0, \Delta}, \pi_D, \mu_{1, \Delta}$ & \cellcolor{orange!40} 0.124 (0.038) & \cellcolor{red!40} 0.232 (0.082) & \cellcolor{orange!40} 0.126 (0.040) \\
$\pi_A, \mu_{0, \Delta}, \pi_D, \mu_{1, \Delta}$ & \cellcolor{red!40} 0.178 (0.066) & \cellcolor{red!40} 0.232 (0.082) & \cellcolor{red!40} 0.165 (0.063) \\
\hline
\end{tabular}
\end{center}
\end{table}

Under correct nuisance function specification, $\hat{\Psi}^{(or)}(\delta)$ demonstrates the lowest bias and highest efficiency. 
The lower performance of $\hat{\Psi}^{(mr)}(\delta)$ in these scenarios is the cost of flexibility by using a non-parametric regression in Step (3) of our estimation procedure, which induces slower convergence and centers around a smoothed version of the effect curve (Web Appendix D).
To this point, we were able to replicate the bias and efficiency of $\hat{\Psi}^{(or)}(\delta)$ using $\hat{\Psi}^{(mr)}(\delta)$ by replacing Step (3) with a parametric regression $\hat{\xi} \sim D + D^3$ (Table A3).
However, a misspecified parametric form will inhibit consistency regardless of nuisance function specification.
The RMSE of $\hat{\Psi}^{(ipw)}(\delta)$, roughly 1.5-2x that of $\hat{\Psi}^{(mr)}(\delta)$ and $\hat{\Psi}^{(or)}(\delta)$ in well-specified scenarios depending on the sample size, demonstrates the instability of approaches that rely solely on weighting even under correct model specification. 
Still, $\hat{\Psi}^{(mr)}(\delta)$ performance may exhibit similar instability under misspecified outcome models.

Confidence intervals for $\hat{\Psi}^{(mr)}$ from both sandwich variance and bootstrap approaches were quite similar and achieved slightly less than the specified coverage probabilities in scenarios with consistent effect estimates for $n=200$ and $n=1000$ sample sizes.
The under-coverage, which worsens in the $n=5000$ simulations, is driven by the finite sample bias of the local linear kernel regression technique.
Demonstrating this fact, the coverage rates of corresponding sandwich and bootstrap confidence intervals for $\hat{\theta}_0$ are between $94.2-96.1\%$ across the sample sizes, whereas those for $\hat{\theta}(\delta)$ appear to converge to $89.5-90.5\%$.
With large samples, the non-parametric convergence rates of $O_p(1/\sqrt{n_Ah} + h^2)$ from the kernel regression dominate the convergence and asymptotic distribution of $\hat{\Psi}(\delta)$, driving the coverage towards that of $\hat{\theta}(\delta)$.
When we correct the finite sample bias by subtracting off the mean bias across simulations, we achieve between $94.9-95.5\%$ coverage, suggesting our intervals have accurate width but imprecise centering. 

Compared to asymptotic variance formulas for doubly robust DiD methods under a binary treatment based on semi-parametric efficiency bounds which are invalid when either nuisance function is incorrectly specified, our flexible variance approaches appear to be more robust to model misspecification in simulations \citep{Santanna2020}.
Still, although previous works have found bootstrap approaches to be robust to model misspecification, this ``multiply robust for inference" property has not been proven theoretically for either bootstrap or sandwich variance approaches \citep{Funk2011DoublyEffects}.

\begin{table}
\caption{A comparison of Coverage Probabilities (Interval Widths) for confidence intervals derived from sandwich and bootstrap variances for the proposed, multiply robust estimator in simulation scenarios varying by nuisance function specification.}
\label{t:sims_coverage}
\begin{center}
\begin{tabular}{lcccccc}
\hline
Incorrect Models & \multicolumn{2}{c}{$n=200$} & \multicolumn{2}{c}{$n=1000$} & \multicolumn{2}{c}{$n=5000$} \\
& Sandwich & Bootstrap & Sandwich & Bootstrap & Sandwich & Bootstrap \\
\hline
None & 93.8 (0.81) & 93.2 (0.78) & 93.8 (0.37) & 93.2 (0.36) & 91.0 (0.17) & 90.7 (0.17) \\
$\pi_A, \pi_D$ & 93.7 (0.81) & 93.4 (0.79) & 93.2 (0.36) & 92.9 (0.36) & 91.3 (0.17) & 90.9 (0.17)  \\
$\mu_{0,\Delta}, \mu_{1,\Delta}$ & 96.1 (1.19) & 93.7 (1.05) & 96.8 (0.54) & 94.0 (0.46) & 97.5 (0.24) & 94.5 (0.21) \\
$\pi_A, \mu_{0, \Delta}, \pi_D, \mu_{1, \Delta}$ & 89.0 (1.18) & 87.3 (1.12) & 70.0 (0.53) & 67.0 (0.51) & 40.5 (0.23) & 38.4 (0.23) \\
\hline
\end{tabular}
\end{center}
\end{table}

\section{Application to Nutritional Excise Taxes}
\label{s:app}

\subsection{Philadelphia Beverage Tax Study}

We applied our proposed methodology to estimate the effects of an excise tax on sugar-sweetened and artificially-sweetened beverages implemented in January 2017 in Philadelphia, Pennsylvania (PA).
Previous studies have largely used difference-in-differences methods to estimate the average effect of the tax on volume sales of taxed beverages in Philadelphia and non-taxed neighboring county stores and estimated large effects in both regions, suggesting cross-border shopping effects \citep{Roberto2019AssociationSetting, Hettinger2023EstimationTax, Cawley2019TheChildren}. 
While these studies primarily assessed the impact of the tax as a binary intervention, they also found stronger tax effects associated with stores closer to the Philadelphia border. 
However, these studies lacked a robust causal framework to effectively control for confounding factors between border proximity and beverage sales.

Here, we analyzed 140 pharmacies from Philadelphia (treated group) and 123 pharmacies from Baltimore and non-neighboring PA counties (control group) to estimate the average effect of the tax on Philadelphia stores were all Philadelphia stores $\delta$ miles from a non-taxed store. 
Data provided by Information Resources Inc. (IRI) and described previously included volume sales and prices of taxed beverages aggregated in each 4-week period in the year prior to (2016) and after (2017) tax implementation ($m=1,...,13$) \citep{MuthIRI2016, Roberto2019AssociationSetting}.
To account for variations in store size, we scaled volume sales for each store by their average 4-week sales in 2016, which serves to both mitigate many differences in outcome trends between stores and also improve the interpretability of findings.
As our continuous measure representing the exposure to cross-border shopping, we calculated the land distance between the centroid of each Philadelphia zip code and the nearest non-taxed zip code.
Accessibility to cross-border shopping could be influenced by expanding taxes geographically or introducing between-region tolls~\citep{Roberto2019AssociationSetting}.
We considered the weighted price of taxed beverages per unit per ounce at each store in 2016 and zip code-level social deprivation index (SDI) information derived from 2012-2016 American Community Survey (ACS) data as confounders~\citep{TheRobertGrahamCenter2018SocialSDI}.
Pre-tax data reveal imbalanced distributions of confounders across intervention and exposure status (Table~\ref{t:rda_t1}).

\begin{table}[b]
\caption{Descriptive statistics for pharmacies in Philadelphia beverage tax study. Mean (standard deviation) metrics are calculated from 2016 data across stores in a given subset.}
\centering
\label{t:rda_t1}
\fbox{%
\begin{tabular}{l|ccc|cc}
 & \multicolumn{3}{c|}{Philadelphia} & \multicolumn{2}{c}{Control}\\[1pt]
 &  $<3$ miles & $3-4.5$ miles & $>4.5$ miles &Baltimore & Non-Border PA \\
 Confounder & $(n=42)$ & $(n=52)$ & $(n=46)$ & $(n=45)$ & $(n=78)$\\
 \hline
Sales (thousand oz.)  & 140.5 (83.7)  & 151.1 (86.3)  & 190.1 (100.2) & 143.6 (62.7)  & 74.1 (44.2)     \\
Price (cents/oz.)         & 6.4 (0.6)     & 6.8 (0.7)     & 6.7 (0.7)     & 6.8 (0.6)     & 7.3 (0.6)     \\
SDI Score          & 83.4 (10.8)   & 66.0 (22.0)       & 77.0 (20.0)       & 77.6 (20.9)   & 17.6 (15.1)   \\
Poverty Score     & 82.4 (11.0)     & 62.6 (24.8)   & 75.6 (22.5)   & 72.6 (23.2)   & 14.3 (12.4)  \\
Education Score    & 71.0 (15.4)     & 54.0 (23.9)     & 64.8 (27.6)   & 64.6 (21.1)   & 23.0 (15.8)       \\
No Vehicle Score    & 93.6 (6.2)    & 91.4 (10.3)   & 93.8 (8.4)    & 92.1 (10.7)   & 37.4 (25.0)       \\
Nonemployed Score  & 91.2 (14.4)   & 61.4 (27)     & 71.5 (28.3)   & 74.4 (25.0)     & 34.9 (18.8)     \\
\end{tabular}}
\end{table}

\subsection{Repeated Observation Times and Nuisance Function Estimation}

Notably, our study differs from those concerning staggered treatment adoption and dynamic treatment regimes as all treated units in our study were intervened on at the same time and exposure was constant for all post-intervention observation times.
To incorporate the repeated observation time structure of our data into our analysis, we followed the approach of \cite{Hettinger2023EstimationTax}, which considered pairs of observation times matched on calendar time ($m$) before and after the tax implementation.
Presuming our parallel trends assumptions now hold between each of the 13 pairs of corresponding observation times, we then calculated $m$-time specific effects using the previously described methods and averaged them over all $m$.
Our described bootstrap methods directly extend to this setting as they account for correlation within subjects over time by re-sampling units, whereas sandwich variance estimators require additional extensions (Web Appendix E). 

We then estimated our nuisance functions separately for each $m$ using SuperLearner, which optimally combines different machine and statistical learning algorithms to limit the potential for model misspecification \citep{vanderLaan2007SuperLearner}.
Specifically, we included a comprehensive ensemble of generalized linear models, generalized additive models, Lasso, and boosting algorithms as candidate learners for the conditional mean functions of our binomial ($\hat{\pi}_A$) and gaussian ($\hat{\pi}_D, \hat{\mu}_{0,\Delta}, \hat{\mu}_{1,\Delta}$) exposures and outcomes. 
To model the conditional density function of the continuous exposure, we additionally modeled the squared residuals from the mean model, $\hat{\epsilon}_i^2 = (Y_{i1}-Y_{i0}-\hat{E}[D | A=1,\mathbf{X_i}])^2$, using SuperLearner and then applied a kernel density estimator for $(D_i - \hat{D}_i)/\sqrt{\hat{\epsilon}_i^2}$ \citep{ Kreif2015EvaluationInjury}.

\subsection{Analysis of Effect Curves}

\begin{figure}
\centering
\includegraphics[width=0.48\textwidth]{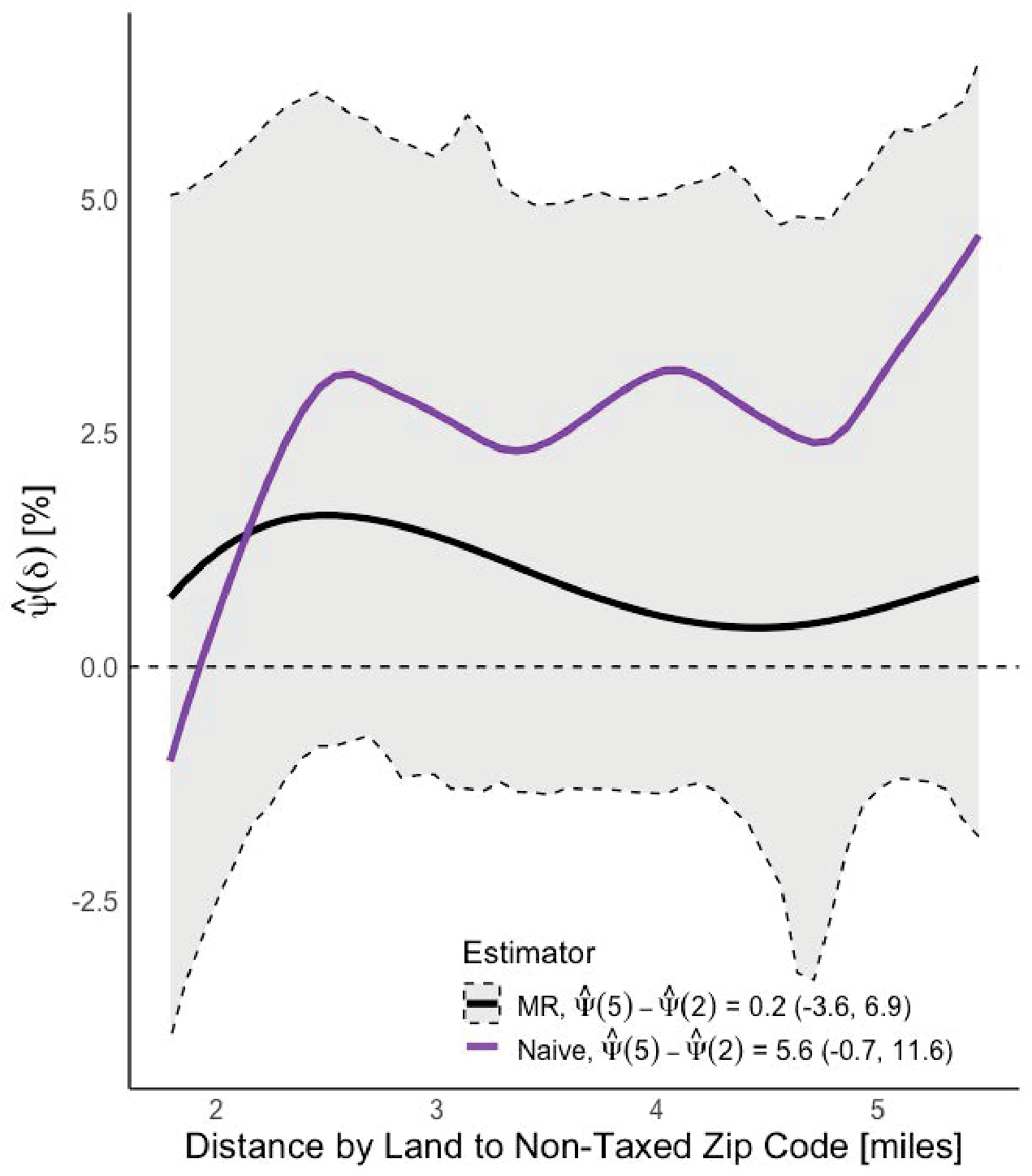}
\includegraphics[width=0.48\textwidth]{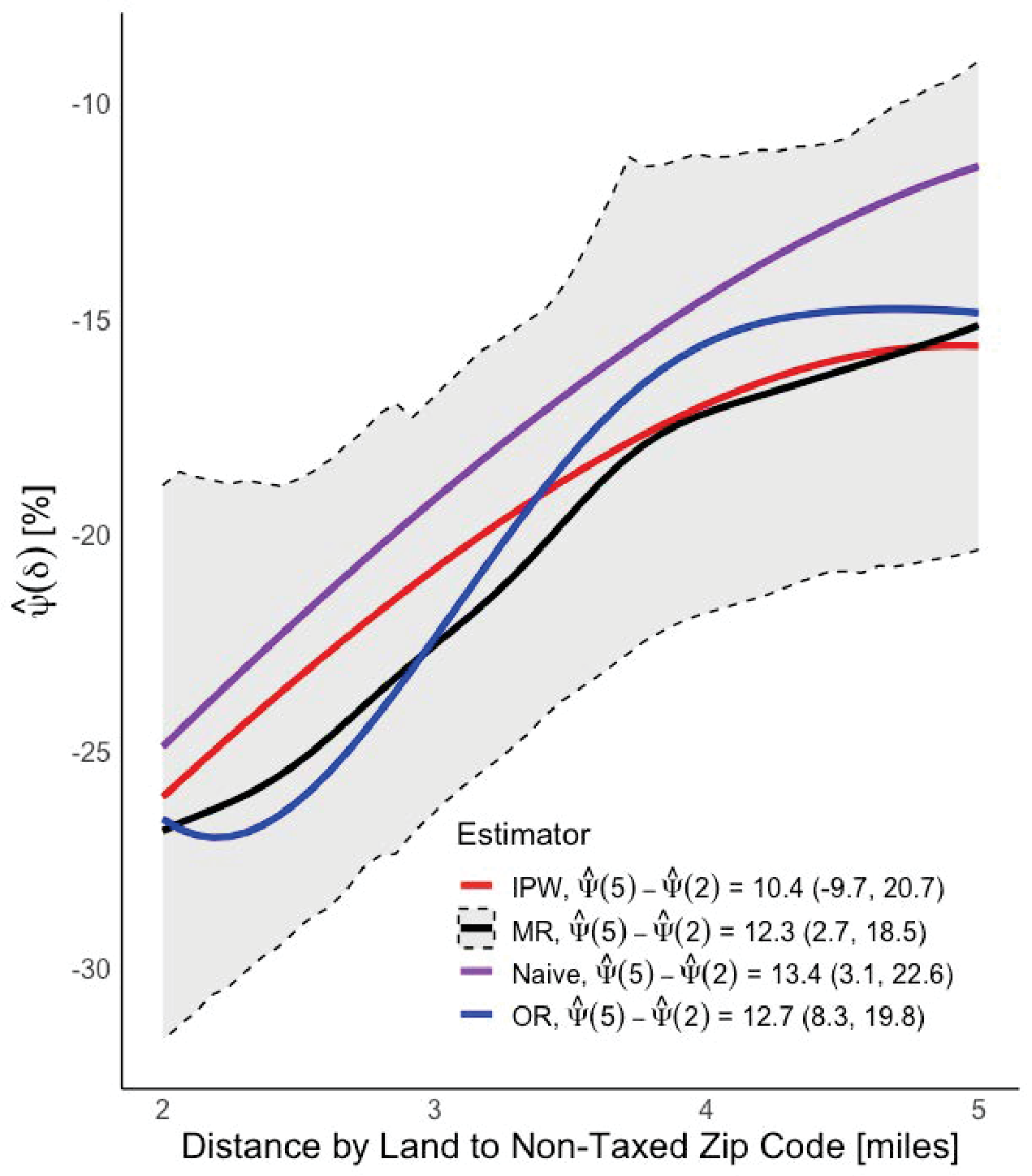}
\caption{Causal effect curve estimates representing the effect of a tax where the nearest non-taxed store for all Philadelphia stores is set at $\delta$ miles away. On the left, pre-trends effect (e.g., placebo test) estimates. On the right, tax effect estimates. In the plot legends, estimates for the difference in tax effects under two distances. Estimates are calculated using the multiply robust (MR), inverse probability weighted (IPW), outcome regression (OR), and confounder-naive (Naive) approaches with pointwise confidence intervals derived for the multiply robust estimator from bootstrap confidence intervals.}
\label{fig:effectcurves}
\end{figure}

We first used our methodology to assess parallel trends in the pre-treatment period as described in Section~\ref{s:methods} by testing known placebo effects for the trends between $m=1$ and $m=2,...,13$ observation times in 2016.
Here, we treated January 2016 as the ``pre-intervention”, $t=0$, time period and the subsequent 2016 periods individually as ``post-intervention”, $t=1$, time periods. 
In addition to only serving as a proxy for the untestable counterfactual parallel trends assumptions, these placebo tests do not directly assess conditional parallel trends in the pre-treatment period as such tests would require estimation of conditional effects (by $\mathbf{X}$), a challenging area of research itself. 
Instead, these placebo tests assess for confounding on the population level after adjusting for potential confounders using our methodology. 
In our setting, average pre-trends effects were less than $5\%$ with either the multiply robust or confounding-naive approaches.
However, estimated placebo effect curves were noticeably smaller, flatter, and with tighter confidence bounds when using our multiply robust method, implying potential benefits from our confounder adjustment (Figure~\ref{fig:effectcurves}). 

We then estimated causal effect curves under the described multiply robust, IPW, outcome regression, and confounder-naive approaches (Figure~\ref{fig:effectcurves}). 
Using our proposed approach, we estimated that a tax policy where all Philadelphia stores were 2 miles (5th percentile of $D$) from the nearest non-taxed zip code would have resulted in a 27.4\% (95\% CI, [18.8, 31.9]) decline in sales, whereas a tax policy where all Philadelphia stores were 5 miles (95th percentile of $D$) from the nearest non-taxed zip code would have resulted in a 15.1\% (95\% CI, [9.0, 20.3]) decline in sales (12.3\% difference, 95\% CI [2.7, 18.5]). 
This substantial difference would indicate that the tax would have a significant effect even under restricted accessibility to cross-border shopping, but that cross-border shopping leads to considerable effect heterogeneity even after adjusting for the spatially-distributed subpopulation characteristics within Philadelphia.
Without adjusting for confounding ($\hat{\Psi}^{(naive)}(\delta)$), we estimated a steeper effect curve, suggesting that economic and demographic differences over distance may explain some of the observed association between border proximity and sales trends.
Outcome regression and IPW approaches estimated similar effects of distance, with the former resembling $\hat{\Psi}^{(mr)}(\delta)$ for $\delta \in (2,3.1)$ and the latter resembling $\hat{\Psi}^{(mr)}(\delta)$ for $\delta \in (3.8,5)$.
Notably, the IPW approach is sensitive to extreme weights even under weight normalization and resulted in wide confidence intervals.

Researchers can use estimates like these to attribute effect heterogeneity to cross-border shopping behaviors after adjusting for economic and demographic differences or determine thresholds at which people will no longer travel to avoid the tax. 
To this end, future studies can improve on these assessments if de-identified stores, rather than zip codes, can be linked to SDI, transportation accessibility, and travel cost data.

\section{Discussion}
\label{s:discuss}

In this paper, we introduced a novel approach for estimating causal effect curves for continuous exposures within the framework of difference-in-differences designs. 
These effect curves enable researchers to explore questions such as, “What would be the average effect of a policy were all units exposed to the policy at the $\delta$-level?”, while accounting for multiple sources of observed confounding commonly encountered in policy evaluations. 
Crucially, our approach offers a flexible and multiply robust mechanism for confounder adjustment without imposing parametric assumptions on the form of the effect curve.
Further, our approach can incorporate general machine learning and non-parametric modeling techniques by either imposing a limited set of restrictions or utilizing sample splitting techniques \citep{Bonvini2022FastEstimation}. 
This flexibility and robustness are particularly vital given the additional modeling complexities in this setting with a continuous exposure and two sources of confounding compared to the standard binary exposure setting. 
Moreover, by harnessing information from both treatment/exposure and outcome models through an influence function-based estimator, our approach enhances the efficiency of existing approaches that rely solely on propensity score weights.

In addition to estimating effect curves, policymakers can use these methods to consider policy effects under counterfactual settings, like when a geographic border exists between taxed and non-taxed regions or neighboring regions are also taxed.
Whereas our approach would rely heavily on representativeness of Philadelphia stores far from the city border for this counterfactual scenario, \cite{Lee2023PolicySpillover} developed methodology to assess such questions by exploiting assumptions about which sale increases in non-taxed neighboring counties would be returned to Philadelphia had neighboring counties also been taxed. 
Incorporating assumptions regarding counterfactual consumer behaviors may also improve our methodology when applied to similar questions.

While designed for difference-in-differences studies like the application presented, the fundamental concepts of our approach have broader applications. 
For example, one can readily extend our methodology to accommodate alternative study designs, such as the controlled interrupted time series, by modifying the presented parallel trends assumptions to the new design and adapting the derived influence function according to the analogous conditional expectations for $\mu_{1,\Delta}$ and $\mu_{0,\Delta}$ \citep{LopezBernal2018TheInterventions}. 
Further, our specification of potential outcomes under a two-dimensional exposure vector resembles the exposure mapping framework introduced by \cite{Aronow2013}. 
In this framework, our continuous exposure, $D$, is analogous to their exposure mapping function, which represents the amount of exposure received through one’s neighbors. 
As a result, our approach holds promise for estimation of effects specified under interference, where current approaches are highly sensitive to confounding \citep{Butts2021}. 
To fully realize these applications, however, future work must address uncertainty estimation when units are dependent and spatially correlated.

\section*{Acknowledgements}

This work was supported by NSF Grant 2149716 (PIs: Mitra and Lee).

\vspace*{-8pt}




\bibliographystyle{unsrtnat}
\bibliography{references}

\appendix

\setcounter{equation}{0}
\setcounter{figure}{0}
\setcounter{table}{0}
\setcounter{section}{0}
\renewcommand{\theequation}{A\arabic{equation}}
\renewcommand{\thefigure}{A\arabic{figure}}
\renewcommand{\thesection}{A\arabic{section}}
\renewcommand{\thetable}{A\arabic{table}}

\section*{Web Appendix A: Proof of Theorem 1}

\begin{align*}
    \Psi(\delta) &= E[Y_1^{(1,\delta)} - Y_1^{(0,\emptyset)} | A=1] \\
    &= E[E[Y_1^{(1,\delta)} - Y_1^{(0,\emptyset)} | A=1, \mathbf{X}] | A=1] \\
    &= E[E[Y_1^{(1,\delta)}-Y_0^{(0,\emptyset)} | A=1, \mathbf{X}] -E[Y_1^{(0,\emptyset)} - Y_0^{(0,\emptyset)} | A=1, \mathbf{X}] | A=1] \\
    &= E[E[Y_1^{(1,\delta)}-Y_0^{(0,\emptyset)} | A=1,  D=\delta, \mathbf{X}] - E[Y_1^{(0,\emptyset)} - Y_0^{(0,\emptyset)} | A=0, \mathbf{X}] | A=1] \\
    &= E[E[Y_1-Y_0 | A=1,  D=\delta, \mathbf{X}]  - E[Y_1 - Y_0 | A=0, \mathbf{X}] | A=1] \\
\end{align*}

The first line comes by definition, the second by iterated expectations, and the third by adding $Y_0^{(0,\emptyset)} - Y_0^{(0,\emptyset)} = 0$. The fourth line comes by invoking (A5) for the right-hand side of the equation and (A6) for the left-hand side. The fifth line holds due to (A1-A3) and is identified with fully observable data. By definition, $\Psi$ is also identifiable under the same assumptions.

\section*{Web Appendix B: Proof of Theorem 2}

Let $p(\mathbf{O};\epsilon)$ be a parametric submodel for the data $\mathbf{O}=(\mathbf{X}, A, D, Y_0, Y_1)$ with parameter $\epsilon \in \mathbbm{R}$ and $p(\mathbf{O};0) = p(\mathbf{O})$. By definition, the efficient influence function for $\Psi$ is the unique function $\phi(\mathbf{O})$ such that $\Psi_\epsilon'(0) = E[\phi(\mathbf{O})\ell_\epsilon'(\mathbf{O};0)]$, where $\Psi(\epsilon)$ represents the parameter of interest as a function on the parametric submodel, $g_\epsilon'(\mathbf{t};0) = \{\partial g(\mathbf{t};\epsilon) \partial \epsilon\}|_{\epsilon=0}$ for any general function $g$ of $\epsilon$ and other arguments $\mathbf{t}$, and $\ell(\mathbf{w}|\mathbf{\Bar{w}}; \epsilon) = \text{log}p(\mathbf{w}|\mathbf{\Bar{w}}; \epsilon)$ for any partition $(\mathbf{W}, \mathbf{\Bar{W}}) \subseteq \mathbf{O}$. 

Then, $p(\mathbf{o}) = \{p(y_1,y_0|A=1, \mathbf{x}, \delta)p(\delta|A=1, \mathbf{x})p(A=1 | \mathbf{x})p(\mathbf{x})\}^a + \{p(y_1,y_0|A=0, \mathbf{x})p(A=0 | \mathbf{x})p(\mathbf{x})\}^{1-a}$ and $\ell_\epsilon'(\mathbf{o};\epsilon) = a[\ell_\epsilon'(y_1,y_0|A=1, \mathbf{x}, \delta;\epsilon) + \ell_\epsilon'(\delta|A=1, \mathbf{x}; \epsilon) + \ell_\epsilon'(A=1 | \mathbf{x}; \epsilon) + \ell_\epsilon'(\mathbf{x};\epsilon)] + (1-a)[\ell_\epsilon'(y_1,y_0|A=0, \mathbf{x};\epsilon) + \ell_\epsilon'(A=0 | \mathbf{x}; \epsilon) + \ell_\epsilon'(\mathbf{x}; \epsilon)]$.

We note the following four properties consistently invoked in our proofs that directly result from properties of log-transformations, score functions, derivatives, and conditional expectations.
\begin{itemize}
    \item Property 1: $\ell_\epsilon'(\mathbf{w} | \mathbf{\Bar{w}}; \epsilon) = p_\epsilon'(\mathbf{w}|\mathbf{\Bar{w}}; \epsilon) / p(\mathbf{w}|\mathbf{\Bar{w}}; \epsilon)$
    \item Property 2: $E[\ell_\epsilon'(\mathbf{W} | \mathbf{\Bar{W}};0) | \mathbf{\Bar{W}}] = 0$
    \item Property 3: $\ell_\epsilon'(\mathbf{w}, \mathbf{\Bar{w}} ; \epsilon) = \ell_\epsilon'(\mathbf{w} | \mathbf{\Bar{w}}; \epsilon) + \ell_\epsilon'(\mathbf{\Bar{w}}; \epsilon) = \ell_\epsilon'(\mathbf{\Bar{w}} | \mathbf{w}; \epsilon) + \ell_\epsilon'(\mathbf{w} ; \epsilon)$
    \item Property 4: $E[B g(\mathbf{O})] = E[P(B=1) g(\mathbf{O}) | B=1]$ for a general function $g$ and a binary random variable $B \in \{0,1\}$
\end{itemize}

By definition, $\Psi(\epsilon) = \int\limits_\mathbbm{D} \Psi(\delta; \epsilon) f(\delta | A=1; \epsilon) d \delta$ and therefore, $\Psi_\epsilon'(0) = \int\limits_\mathbbm{D} \{\Psi_\epsilon'(\delta; 0) f(\delta | A=1) + \Psi(\delta) f_\epsilon'(\delta | A=1;0) \} d\delta = E[\Psi_\epsilon'(D; 0) + \Psi(D) l_\epsilon'(D | A=1;0) | A=1]$ by Property 1. The second term can be simplified as $E[\Psi(D) l_\epsilon'(D | A=1;0) | A=1] = E[\theta(D) l_\epsilon'(D | A=1;0) | A=1] - \theta_0 E[l_\epsilon'(D | A=1;0) | A=1] = E[\theta(D) l_\epsilon'(D | A=1;0) | A=1]$, where the first equality holds by definition of $\Psi(D)$ and the second by Property 2. As $\Psi(\delta; \epsilon) = \int\limits_\mathbbm{X} \int\limits_\mathbbm{Y} (y_1-y_0) [p(y_1, y_0 | A=1, \mathbf{x}, \delta; \epsilon) - p(y_1, y_0 | A=0, \mathbf{x}] p(\mathbf{x}|A=1;\epsilon) d(y_1, y_0) d\mathbf{x}$, we have, $\Psi_\epsilon'(\delta;0)$ 

\begin{align*}
     &= \int\limits_\mathbbm{X} \int\limits_\mathbbm{Y} (y_1-y_0) \{p_\epsilon'(y_1,y_0 | A=1, \mathbf{x}, \delta; 0) p(\mathbf{x}|A=1) + p(y_1,y_0|A=1,\mathbf{x}, \delta)p_\epsilon'(\mathbf{x}|A=1;0)\} d(y_1,y_0) d\mathbf{x} \\
     & \qquad - \int\limits_\mathbbm{X} \int\limits_\mathbbm{Y} (y_1-y_0) \{p_\epsilon'(y_1,y_0 | A=0, \mathbf{x}; 0) p(\mathbf{x}|A=1) + p(y_1,y_0|A=0,\mathbf{x})p_\epsilon'(\mathbf{x}|A=1;0)\}d(y_1,y_0) d\mathbf{x} \\
    &= \int\limits_\mathbbm{X} \int\limits_\mathbbm{Y} (y_1-y_0) \{\ell_\epsilon'(y_1,y_0 | A=1, \mathbf{x}, \delta; 0)p(y_1,y_0|A=1,\mathbf{x}, \delta) p(\mathbf{x}|A=1) + \\
    & \qquad p(y_1,y_0|A=1,\mathbf{x}, \delta)\ell_\epsilon'(\mathbf{x}|A=1;0) p(\mathbf{x}|A=1)\}d(y_1,y_0) d\mathbf{x} - \\
    & \qquad \int\limits_\mathbbm{X} \int\limits_\mathbbm{Y} (y_1-y_0) \{\ell_\epsilon'(y_1,y_0 | A=0, \mathbf{x}; 0)p(y_1,y_0|A=0,\mathbf{x}) p(\mathbf{x}|A=1) + \\
    & \qquad p(y_1,y_0|A=0,\mathbf{x})\ell_\epsilon'(\mathbf{x}|A=1;0) p(\mathbf{x}|A=1)\}d(y_1,y_0) d\mathbf{x} \\
    &= E[E[(Y_1-Y_0)\ell_\epsilon'(Y_1,Y_0 | A=1, \mathbf{X}, D=\delta; 0)|A=1,\mathbf{X},D=\delta] | A=1] + \\
    & \qquad E[\mu_{1,\Delta}(\mathbf{X}, \delta)\ell_\epsilon'(\mathbf{X}|A=1;0)|A=1]  - \\
    & \qquad E[E[(Y_1-Y_0) \ell_\epsilon'(Y_1,Y_0 | A=0, \mathbf{X}; 0) | A=0,\mathbf{X})] | A=1] - E[\mu_{0,\Delta}(\mathbf{X}) \ell_\epsilon'(\mathbf{x}|A=1;0) |A=1)]
\end{align*}

Here, the first equality comes by definition, the second from use of Property 1, and the third by re-writing as expectations. Therefore, our first target quantity is,
\begin{align*}
    \Psi_\epsilon'(0) &= \int\limits_\mathbbm{D} \{ E[E[(Y_1-Y_0)\ell_\epsilon'(Y_1,Y_0 | A=1, \mathbf{X}, D=\delta; 0)|A=1,\mathbf{X},D=\delta] | A=1] + \\
    & \qquad E[\mu_{1,\Delta}(\mathbf{X}, \delta)\ell_\epsilon'(\mathbf{X}|A=1;0)|A=1]  - E[\mu_{0,\Delta}(\mathbf{X}) \ell_\epsilon'(\mathbf{x}|A=1;0) |A=1)] + \\
    & \qquad \theta(\delta) l_\epsilon'(\delta | A=1;0) - E[E[(Y_1-Y_0) \ell_\epsilon'(Y_1,Y_0 | A=0, \mathbf{X}; 0) | A=0,\mathbf{X})] | A=1]  \} df(\delta | A=1)
\end{align*}

We then demonstrate the equivalence between $\Psi_\epsilon'(0)$ and the covariance, $E[\phi(\mathbf{O})\ell_\epsilon'(\mathbf{O};0)] = E[\phi(\mathbf{O})\{A[\ell_\epsilon'(y_1,y_0|A=1, \mathbf{X}, D;\epsilon) + \ell_\epsilon'(D|A=1, \mathbf{X}; \epsilon) + \ell_\epsilon'(A=1 | \mathbf{X}; \epsilon) + \ell_\epsilon'(\mathbf{X};\epsilon)] + (1-A)[\ell_\epsilon'(y_1,y_0|A=0, \mathbf{X};\epsilon) + \ell_\epsilon'(A=0 | \mathbf{X}; \epsilon) + \ell_\epsilon'(\mathbf{X}; \epsilon)]\}]$ under our previously specified influence function, $\phi(\mathbf{O}) = \frac{A}{P(A=1)}[\frac{f(D|A=1)}{\pi_D(D|A=1,\mathbf{X})}(Y_1-Y_0 - \mu_{1,\Delta}(\mathbf{X}, D)) + m(D|A=1)] - \frac{A}{P(A=1)}\mu_{0,\Delta}(\mathbf{X}) - \frac{1-A}{P(A=1)}\frac{\pi(\mathbf{X})}{1-\pi(\mathbf{X})}(Y_1-Y_0-\mu_{0,\Delta}(\mathbf{X})) + \frac{A}{P(A=1)}\int\limits_\mathbbm{D} \{\mu_{1,\Delta}(\mathbf{X}, \delta) - m(\delta | A=1) \} df(\delta|A=1) - \Psi$.

Using Property 4 and Property 2 in conjunction with iterated expectations conditioning on $\mathbf{X}, D,$ and $A=1$, we can rearrange the first term as $E[\phi(\mathbf{O})A\ell_\epsilon'(y_1,y_0|A=1, \mathbf{X}, D;\epsilon)]$
\begin{align*}
     &= E[\frac{A}{P(A=1)} \{\frac{f(D|A=1)}{\pi_D(D|A=1,\mathbf{X})}(Y_1-Y_0-\mu_{1,\Delta}(\mathbf{X}, D)) + m(D|A=1) - \mu_{0,\Delta}(X) \\
     & \qquad + \int\limits_\mathbbm{D} (\mu_{1,\Delta}(\mathbf{X},\delta) - m(\delta|A=1))df(d|A=1) - P(A=1)\Psi\} \ell_\epsilon'(Y_1,Y_0|A=1,\mathbf{X},D;0)]\\
    &= E[\frac{f(D|A=1)}{\pi_D(D|A=1,\mathbf{X})}(Y_1-Y_0-\mu_{1,\Delta}(\mathbf{X}, D)) \ell_\epsilon'(Y_1,Y_0|A=1,\mathbf{X},D;0) | A=1]\\
    &= \int\limits_\mathbbm{D} \int\limits_\mathbbm{X} \int\limits_\mathbbm{Y} (y_1-y_0) \ell_\epsilon'(y_1,y_0|A=1,\mathbf{x},\delta;0) dP(y_1,y_0|A=1,\mathbf{x},\delta)  \\
     & \qquad \{ \frac{f(\delta|A=1)}{\pi_D(\delta|A=1,\mathbf{X})} dP(\mathbf{x}|A=1, \delta) \} df(\delta|A=1) \\
    &= \int\limits_\mathbbm{D} \int\limits_\mathbbm{X} \int\limits_\mathbbm{Y} (y_1-y_0) \ell_\epsilon'(y_1,y_0|A=1,\mathbf{x},\delta;0) dP(y_1,y_0|A=1,\mathbf{x},\delta) dP(\mathbf{x}|A=1) df(\delta|A=1) \\
    &= \int\limits_\mathbbm{D} E[E[ (Y_1-Y_0) \ell_\epsilon'(Y_1,Y_0|A=1,\mathbf{X},\delta;0) |A=1,\mathbf{X},\delta)] | A=1] df(\delta|A=1)
\end{align*}

Similarly, we can break down the second term as
\begin{align*}
    & E[\phi(\mathbf{O})A\ell_\epsilon'(D, \mathbf{X}|A=1;\epsilon)] = E[\{\frac{f(D|A=1)}{\pi_D(D|A=1,\mathbf{X})}(Y_1-Y_0-\mu_{1,\Delta}(\mathbf{X}, D)) + m(D|A=1) \\
    & \qquad - \mu_{0,\Delta}(X) + \int\limits_\mathbbm{D} (\mu_{1,\Delta}(\mathbf{X},\delta) - m(\delta|A=1))df(d|A=1) - P(A=1)\Psi\} \ell_\epsilon'(D, \mathbf{X}|A=1;\epsilon) | A=1]\\
    & E[\{\frac{f(D|A=1)}{\pi_D(D|A=1,\mathbf{X})}(Y_1-Y_0-\mu_{1,\Delta}(\mathbf{X}, D)) \} \ell_\epsilon'(D, \mathbf{X}|A=1;\epsilon) | A=1] = 0 \\
    & E[\{m(D|A=1) \} \ell_\epsilon'(D, \mathbf{X}|A=1;\epsilon) | A=1] = E[\theta(D) \ell_\epsilon'(D, \mathbf{X}|A=1;\epsilon) | A=1] \\
    & E[\{\mu_{0,\Delta}(X) \} \ell_\epsilon'(D, \mathbf{X}|A=1;\epsilon) | A=1] = E[\mu_{0,\Delta}(X) \ell_\epsilon'(D | \mathbf{X}, A=1;\epsilon) | A=1] \\
    & \qquad + E[\mu_{0,\Delta}(X) \ell_\epsilon'(\mathbf{X} | A=1;\epsilon) | A=1] \\
    & \qquad = E[\mu_{0,\Delta}(X) \ell_\epsilon'(\mathbf{X} | A=1;\epsilon) | A=1] \\
    & E[\ell_\epsilon'(D, \mathbf{X}|A=1;\epsilon) \{\int\limits_\mathbbm{D} (\mu_{1,\Delta}(\mathbf{X},\delta) - m(\delta|A=1))df(\delta|A=1) \}  | A=1]  \\
    & \qquad = E[\ell_\epsilon'(D |\mathbf{X}, A=1;\epsilon) \int\limits_\mathbbm{D} \mu_{1,\Delta}(\mathbf{X},\delta) df(\delta|A=1)  | A=1] \\
    & \qquad + E[\ell_\epsilon'(\mathbf{X} | A=1;\epsilon) \int\limits_\mathbbm{D} \mu_{1,\Delta}(\mathbf{X},\delta) df(\delta|A=1) | A=1] \\
    & \qquad = E[\ell_\epsilon'(\mathbf{X} | A=1;\epsilon) \int\limits_\mathbbm{D}  \mu_{1,\Delta}(\mathbf{X},\delta) df(\delta|A=1) | A=1]
\end{align*}

Here, the first equation used Property 1, the second equation used iterated expectations conditioning on $\mathbf{X}, D, A=1$ to zero out the $(Y_1-Y_0-\mu_{1,\Delta}(\mathbf{X},D))$ interior, the third equation used the definition of $m(D|A=1)$, the fourth invoked Property 3 and then used Property 2 in conjunction with iterated expectations conditioning on $\mathbf{X}, A=1$, and the fifth used Property 2 and Property 3 followed by Property 2 in conjunction with iterated expectations conditioning on $\mathbf{X}, A=1$.

Shifting to contributions from the control units, $E[\phi(\mathbf{O})(1-A)\ell_\epsilon'(y_1,y_0|A=0, \mathbf{X};\epsilon)]$
\begin{align*}
    &= -E[\frac{P(A=0)}{P(A=1)} \{\frac{\pi_A(\mathbf{X})}{1-\pi_A(\mathbf{X})}(Y_1-Y_0-\mu_{0,\Delta}(\mathbf{X})) - P(A=1) \Psi\} \ell_\epsilon'(Y_1,Y_0|A=0,\mathbf{X};0) | A=0]\\
    &= -\int\limits_\mathbbm{X} \int\limits_\mathbbm{Y} (y_1-y_0 - \mu_{0,\Delta}(\mathbf{x})) \ell_\epsilon'(y_1,y_0|A=0,\mathbf{x};0) dP(y_1,y_0|A=0, \mathbf{x}) \\
    & \qquad \{\frac{P(A=0)}{P(A=1)} \frac{\pi_A(\mathbf{x})}{1-\pi_A(\mathbf{x})} dP(\mathbf{x} | A=0) \}\\
    &= -\int\limits_\mathbbm{X} \int\limits_\mathbbm{Y} (y_1-y_0 - \mu_{0,\Delta}(\mathbf{x}) \ell_\epsilon'(y_1,y_0|A=0,\mathbf{x};0) dP(y_1,y_0|A=0, \mathbf{x}) dP(\mathbf{x} | A=1)\\
    &= -E[E[ (Y_1-Y_0 - \mu_{0,\Delta}(\mathbf{X}) \ell_\epsilon'(Y_1,Y_0|A=0,\mathbf{X};0) | A=0, \mathbf{X}] | A=1]\\
    &= -E[E[ (Y_1-Y_0) \ell_\epsilon'(Y_1,Y_0|A=0,\mathbf{X};0) | A=0, \mathbf{X}] | A=1]
\end{align*}

Here, we used Property 4 to get the first line, the integral definition of expectations in the second and fourth lines, some re-arranging of probability functions in the third line, and Property 2 in conjunction with iterated expectations conditioning on $\mathbf{X}$ and $A=0$  in the fifth line. Then, by using Property 2 followed by iterated expectations conditioning on $\mathbf{X}$ and $A=0$, we get $E[\phi(\mathbf{O})A\ell_\epsilon'(\mathbf{X}|A=0;\epsilon)]$ 
\begin{align*}
     &= -E[\frac{P(A=0)}{P(A=1)}\{\frac{\pi_A(\mathbf{X})}{1-\pi_A(\mathbf{X})}(Y_1-Y_0-\mu_{0,\Delta}(\mathbf{X})) - P(A=1)\Psi\} \ell_\epsilon'( \mathbf{X}|A=0;\epsilon) | A=0]\\
    &= -E[\frac{P(A=0)}{P(A=1)}\{\frac{\pi_A(\mathbf{X})}{1-\pi_A(\mathbf{X})}E[(Y_1-Y_0-\mu_{0,\Delta}(\mathbf{X})) | A=0,\mathbf{X}]\} \ell_\epsilon'( \mathbf{X}|A=0;\epsilon) | A=0]\\
   &= 0
\end{align*}

The final terms $E[\phi(\mathbf{O})A\ell_\epsilon'(A=1;\epsilon)]$ and $E[\phi(\mathbf{O})(1-A)\ell_\epsilon'(A=0;\epsilon)]$ are both zero because, as seen previously, the respective expectations can be written as conditional on $A=1$ or $A=0$, allowing us to readily invoke Property 2 on all terms. Then, $E[\phi(\mathbf{O})\ell_\epsilon'(\mathbf{O};0)]$
\begin{align*}
    &= \int\limits_\mathbbm{D} E[E[ (Y_1-Y_0) \ell_\epsilon'(Y_1,Y_0|A=1,\mathbf{X},\delta;0) |A=1,\mathbf{X},\delta)] | A=1] df(\delta|A=1) \\
    & \qquad + E[\theta(D) \ell_\epsilon'(D, \mathbf{X}|A=1;\epsilon) | A=1] - E[\mu_{0,\Delta}(X) \ell_\epsilon'(\mathbf{X} | A=1;\epsilon) | A=1] \\
    & \qquad + E[\ell_\epsilon'(\mathbf{X} | A=1;\epsilon) \int\limits_\mathbbm{D} \mu_{1,\Delta}(\mathbf{X},\delta) df(\delta|A=1) | A=1] \\
    & \qquad - E[E[ (Y_1-Y_0) \ell_\epsilon'(Y_1,Y_0|A=0,\mathbf{X};0) | A=0, \mathbf{X}] | A=1] \\
    &=\int\limits_\mathbbm{D} \{ E[E[(Y_1-Y_0)\ell_\epsilon'(Y_1,Y_0 | A=1, \mathbf{X}, D=\delta; 0)|A=1,\mathbf{X},D=\delta] | A=1] \\
    & \qquad + \theta(\delta) l_\epsilon'(\delta | A=1;0) - E[\mu_{0,\Delta}(\mathbf{X}) \ell_\epsilon'(\mathbf{x}|A=1;0) |A=1)] \\
    & \qquad + E[\mu_{1,\Delta}(\mathbf{X}, \delta)\ell_\epsilon'(\mathbf{X}|A=1;0)|A=1] \\
    & \qquad  - E[E[(Y_1-Y_0) \ell_\epsilon'(Y_1,Y_0 | A=0, \mathbf{X}; 0) | A=0,\mathbf{X})] | A=1]  \} df(\delta | A=1) \\
    & = \Psi_\epsilon'(0)
\end{align*} 
Thus, $\phi$ is the efficient influence function for $\Psi$.

\section*{Web Appendix C: Proof of Theorem 3}

Recall our defined aggregate parameter, $\Psi = \int\limits_\mathbbm{D} E[\mu_{1,\Delta}(\mathbf{X}, \delta) - \mu_{0,\Delta}(\mathbf{X}) | A=1] df(\delta|A=1)$, and our proposed estimator, $\hat{\Psi} = E_n[\frac{A}{P(A=1)}\hat{\xi}-\frac{A}{P(A=1)}\hat{\mu}_{0,\Delta}(\mathbf{X}) - \frac{1-A}{P(A=1)} \frac{\hat{\pi}_A(\mathbf{X})}{1-\hat{\pi}_A(\mathbf{X})}(Y_1-Y_0-\hat{\mu}_{0,\Delta}(\mathbf{X}))]$. Further, we can consider the decomposition of our parameter as $\Psi = \theta_D - \theta_0$, where $\theta_D = \int\limits_\mathbbm{D} \theta(\delta) df(\delta|A=1) = \int\limits_\mathbbm{D} E[\mu_{1,\Delta}(\mathbf{X}, \delta) | A=1] df(\delta|A=1)$ and $\theta_0 = E[\mu_{0,\Delta}(\mathbf{X}) | A=1] = \int\limits_\mathbbm{D} E[\mu_{0,\Delta}(\mathbf{X}) | A=1] df(\delta|A=1)$. The analogous decomposition of our estimator then is $\hat{\Psi} = \hat{\theta}_D - \hat{\theta}_0$, where $\hat{\theta}_D = E_n[\frac{A}{P(A=1)}\hat{\xi}]$ and $\hat{\theta}_0 = E_n[\frac{A}{P(A=1)}\hat{\mu}_{0,\Delta}(\mathbf{X}) + \frac{1-A}{P(A=1)} \frac{\hat{\pi}_A(\mathbf{X})}{1-\hat{\pi}_A(\mathbf{X})}(Y_1-Y_0-\hat{\mu}_{0,\Delta}(\mathbf{X}))]$.

Let $\Bar{\mu}_{1,\Delta}$, $\Bar{\pi}_D$, $\Bar{\mu}_{0,\Delta}$, and $\Bar{\pi}_A$ be such that $\hat{\mu}_{1,\Delta} \xrightarrow{p} \Bar{\mu}_{1,\Delta}$, $\hat{\pi}_D \xrightarrow{p} \Bar{\pi}_D$, $\hat{\mu}_{0,\Delta} \xrightarrow{p} \Bar{\mu}_{0,\Delta}$, and $\hat{\pi}_A \xrightarrow{p} \Bar{\pi}_A$. Recalling that $\hat{f}(D|A=1) = E_n[\hat{\pi}_D(D | A=1, \mathbf{X}) | A=1]$ and $\hat{m}(D|A=1) = E_n[\hat{\mu}_{1,\Delta}(\mathbf{X}, D) | A=1]$, we will also denote $\Bar{f}(D|A=1) = E_n[\Bar{\pi}_D(D | A=1, \mathbf{X}) | A=1]$ and $\Bar{m}(D|A=1) = E_n[\Bar{\mu}_{1,\Delta}(\mathbf{X}, D) | A=1]$. 

We first consider $\theta_0$ and $\hat{\theta}_0$. Using the law of large numbers and some re-arranging, we see that $\hat{\theta}_0$
\begin{align*}
    &= E_n[\frac{A}{P(A=1)}\hat{\mu}_{0,\Delta}(\mathbf{X}) + \frac{1-A}{P(A=1)} \frac{\hat{\pi}_A(\mathbf{X})}{1-\hat{\pi}_A(\mathbf{X})}] \\
    & \xrightarrow{p} E[\Bar{\mu}_{0,\Delta}(\mathbf{X}) | A=1] + E_n[\frac{P(A=0)}{P(A=1)} \frac{\Bar{\pi}_A(\mathbf{X})}{1-\Bar{\pi}_A(\mathbf{X})} (Y_1-Y_0-\Bar{\mu}_{0,\Delta}(\mathbf{X})) | A=0] \\
    &= E[\Bar{\mu}_{0,\Delta}(\mathbf{X}) | A=1] \\
    & \qquad + \int\limits_\mathbbm{X} \int\limits_\mathbbm{Y} (y_1-y_0-\Bar{\mu}_{0,\Delta}(\mathbf{x})) dP(y_1,y_0|A=0,\mathbf{x}) \frac{P(A=0)}{P(A=1)} \frac{\Bar{\pi}_A(\mathbf{X})}{1-\Bar{\pi}_A(\mathbf{X})} dP(\mathbf{x}| A=0) \\
    & = E[\Bar{\mu}_{0,\Delta}(\mathbf{X}) | A=1] \\
    & \qquad + \int\limits_\mathbbm{X} \int\limits_\mathbbm{Y} (y_1-y_0-\Bar{\mu}_{0,\Delta}(\mathbf{x})) dP(y_1,y_0|A=0,\mathbf{x}) (\frac{\Bar{\pi}_A(\mathbf{X})}{1-\Bar{\pi}_A(\mathbf{X})}) / (\frac{\pi_A(\mathbf{X})}{1-\pi_A(\mathbf{X})}) dP(\mathbf{x}| A=1) \\
    & = E[\Bar{\mu}_{0,\Delta}(\mathbf{X}) + (Y_1-Y_0-\Bar{\mu}_{0,\Delta}(\mathbf{X}))(\frac{\Bar{\pi}_A(\mathbf{X})}{1-\Bar{\pi}_A(\mathbf{X})}) / (\frac{\pi_A(\mathbf{X})}{1-\pi_A(\mathbf{X})})|A=1] \\
    & = E[\mu_{0,\Delta}(\mathbf{X}) + (\Bar{\mu}_{0,\Delta}(\mathbf{X})-\Bar{\mu}_{0,\Delta}(\mathbf{X}))\{(\frac{\Bar{\pi}_A(\mathbf{X})}{1-\Bar{\pi}_A(\mathbf{X})}) / (\frac{\pi_A(\mathbf{X})}{1-\pi_A(\mathbf{X})})-1\} | A=1]
\end{align*}
Therefore, $\hat{\theta}_0 \xrightarrow{p} \theta_0$ if $\Bar{\mu}_{0,\Delta} = \mu_{0,\Delta}$ or $\Bar{\pi}_A = \pi_A$.

We then consider $\theta_D$ and $\hat{\theta}_D$. Similarly, under the law of large numbers and some re-arranging, we see that $\hat{\theta}_D$ 
\begin{align*}
    &= E_n[\frac{A}{P(A=1)}[\frac{\hat{f}(D|A=1)}{\hat{\pi}_D(D|A=1,\mathbf{X})}(Y_1-Y_0-\hat{\mu}_{1,\Delta}(\mathbf{X}, D)) + \hat{m}(D|A=1)]] \\
    & \xrightarrow{p} E[\frac{\Bar{f}(D|A=1)}{\Bar{\pi}_D(D|A=1,\mathbf{X})}(Y_1-Y_0-\Bar{\mu}_{1,\Delta}(\mathbf{X}, D)) + \Bar{m}(D|A=1) | A=1] \\
    & = \int\limits_\mathbbm{D} E[E[\frac{\Bar{f}(\delta|A=1)}{\Bar{\pi}_D(\delta|A=1,\mathbf{X})}(Y_1-Y_0-\Bar{\mu}_{1,\Delta}(\mathbf{X}, \delta)) + \Bar{m}(\delta|A=1) | D=\delta, A=1] | A=1] df(\delta|A=1) \\
\end{align*}
Looking at the interior expectation, $E[\frac{\Bar{f}(\delta|A=1)}{\Bar{\pi}_D(\delta|A=1,\mathbf{X})}(Y_1-Y_0-\Bar{\mu}_{1,\Delta}(\mathbf{X}, \delta)) + \Bar{m}(\delta|A=1) | D=\delta, A=1]$
\begin{align*}
    &= \int\limits_\mathbbm{X} (\mu_{1,\Delta}(\mathbf{x}, \delta) - \Bar{\mu}_{1,\Delta}(\mathbf{x}, \delta)) \frac{\Bar{f}(D|A=1)}{\Bar{\pi}_D(D|A=1,\mathbf{X})} dP(\mathbf{x}|A=1,\delta) + \Bar{m}(\delta|A=1) \\
    & = \int\limits_\mathbbm{X} (\mu_{1,\Delta}(\mathbf{x}, \delta) - \Bar{\mu}_{1,\Delta}(\mathbf{x}, \delta)) (\frac{\Bar{f}(D|A=1)}{\Bar{\pi}_D(D|A=1,\mathbf{X})}) / (\frac{f(D|A=1)}{\pi_D(D|A=1,\mathbf{X})}) dP(\mathbf{x}|A=1) + \Bar{m}(\delta|A=1) \\
    & = E[\mu_{1,\Delta}(\mathbf{X},\delta) | A=1] \\
    & \qquad + \int\limits_\mathbbm{X} (\mu_{1,\Delta}(\mathbf{x}, \delta) - \Bar{\mu}_{1,\Delta}(\mathbf{x}, \delta)) \{ (\frac{\Bar{f}(D|A=1)}{\Bar{\pi}_D(D|A=1,\mathbf{X})}) / (\frac{f(D|A=1)}{\pi_D(D|A=1,\mathbf{X})}) - 1 \} dP(\mathbf{x}|A=1)
\end{align*}
Therefore, $\hat{\theta}_D \xrightarrow{p} \theta_D$ if $\Bar{\mu}_{1,\Delta} = \mu_{1,\Delta}$ or $\Bar{\pi}_D = \pi_D$. Then, for $\hat{\Psi} \xrightarrow{p} \Psi$, we need both $\hat{\theta}_0 \xrightarrow{p} \theta_0$ and $\hat{\theta}_D \xrightarrow{p} \theta_D$, meaning $\hat{\Psi} \xrightarrow{p} \Psi$ if both (i)  $\Bar{\mu}_{0,\Delta} = \mu_{0,\Delta}$ or $\Bar{\pi}_A=\pi_A$ and (ii) $\Bar{\mu}_{1,\Delta} = \mu_{1,\Delta}$ or $\Bar{\pi}_D=\pi_D$.

While the same logic would appear to suffice for $\hat{\Psi}(\delta)$, our use of a non-parametric regressor, rather than a sample mean, for $E[\hat{xi}|A=1, D=\delta]$ precludes our previous invocation of the law of large numbers. Denoting $h$ as the bandwidth parameter of the local linear kernel regression, $\sup_{t: |t-a|\leq h} ||\hat{\pi}_D(t|A=1,\mathbf{X}) - \pi_D(t|A=1,\mathbf{X})|| = O_p(r_{n_A}(\delta))$, $\sup_{t: |t-a|\leq h} ||\hat{\mu}_{1,\Delta}(\mathbf{X},t) - \mu_{1,\Delta}(\mathbf{X},t)|| = O_p(s_{n_A}(\delta))$, $\text{sup} ||\hat{\pi}_A(\mathbf{X}) - \pi_A(\mathbf{X})|| = O_p(q_{n}$, and $\text{sup} ||\hat{\mu}_{0,\Delta}(\mathbf{X}) - \mu_{0,\Delta}(\mathbf{X})|| = O_p(s_{n_A}(\delta))$, we then have
\begin{align*}
    |\hat{\Psi}(\delta) - \Psi(\delta)| &= |\hat{\theta}(\delta) - \theta(\delta) - (\hat{\theta}_0 - \theta_0)| \\
    & \leq |\hat{\theta}(\delta) - \theta(\delta)| + |(\hat{\theta}_0 - \theta_0)| \\
    & = O_p(\frac{1}{\sqrt{n_A h}} + h^2 + r_{n_A}(\delta)s_{n_A}(\delta) + q_n v_{n_0})
\end{align*}
The convergence rate for $\hat{\theta}(\delta)$, $O_p(\frac{1}{\sqrt{n_A h}} + h^2 + r_{n_A}(\delta)s_{n_A}(\delta))$, comes from \cite{Kennedy2017NonparametricEffects}, where our sample only contains the $n_A$ treated units. The last product, $O_p(q_n v_{n_0})$, comes from the convergence rate of the binary ATT \citep{Kennedy2016SemiparametricInferenceb}.

\section*{Web Appendix D: Convergence and Asymptotic Normality}

The inhibiting factor in convergence of absolute bias for $\hat{\Psi}(\delta)$ is generally the slow-convergence of the non-parametric regression, $O_p(\frac{1}{\sqrt{n_A h}} + h^2)$, a cost worth paying to avoid parametric assumptions on the form of the effect curve. Minimizing this term requires $h \sim n_A^{-1/5}$, which results in a convergence rate of $O_p(n_A^{-2/5})$. Then, as long as $O_p(r_{n_A}(\delta)s_{n_A}(\delta)) = O_p(n_A^{-2/5})$ and $O_p(q_n v_{n_0}) = O_p(n_A^{-2/5})$, the nuisance function estimation does not affect convergence rates. These rates are generally slower than the conditions assumed in binary doubly robust estimation, where both nuisance functions are assumed to converge at a rate of $O_p(n^{-1/4})$, giving each product of convergence a rate of $O_p(n^{-1/2}) < O_p(n^{-2/5})$. 

When considering asymptotic normality of $\hat{\Psi}(\delta)$, \cite{Kennedy2017NonparametricEffects} demonstrated asymptotic normality for $\hat{\theta}(\delta)$ if $r_{n_A}(\delta) s_{n_A}(\delta) = O_p(1/\sqrt{n_A h})$. Therefore, we have asymptotic normality of $\hat{\Psi}(\delta)$ under assumptions on the convergence rates of $\hat{\pi}_A$ and $\hat{\mu}_{0,\Delta}$. For example, if we make the standard assumptions of $O_p(n^{-1/4})$ convergence for each function, then $\hat{\theta}_0$ makes no contribution to the limiting distribution of the bias and we have the same asymptotic distribution as \cite{Kennedy2017NonparametricEffects} by Slutsky's theorem. If we were to impose highly specific assumptions such that $O_p(q_n v_{n_0}) = O_p(1/\sqrt{n_A h})$, it would be possible to show some contribution of $\hat{\theta}_0$ in the limiting distribution. However, these distributions would seem inefficient exercises as the lack of knowledge on convergence rates in practice would inhibit their use. Rather, we instead consider the presented sandwich and bootstrap variance estimators.

\section*{Web Appendix E: Sandwich Variance Extensions}

The general form of a sandwich variance estimator is given as $$\boldsymbol{\Gamma_i} = \begin{bmatrix}
EE_{1i} \\
EE_{2i} \\
\vdots \\
EE_{Pi} 
\end{bmatrix},$$
where $EE_{pi}$ represent the $p=1,...,P$ unbiased estimating equations simultaneously solved such that $\sum\limits_{i=1}^n \boldsymbol{\Gamma_i} = \mathbf{0}$. The covariance matrix $V$ is then estimated by $$
    \hat{V}(\hat{\boldsymbol{\eta}}) = \{ \sum\limits_{i=1}^n \frac{\partial}{\partial \eta'} \boldsymbol{\Gamma_i}(\hat{\eta}) \}^{-1} \{ \sum\limits_{i=1}^n \boldsymbol{\Gamma_i}^{\bigotimes 2}(\hat{\eta}) \{ \sum\limits_{i=1}^n \frac{\partial}{\partial \eta} \boldsymbol{\Gamma_i}'(\hat{\eta}) \}^{-1},$$
where $\boldsymbol{\eta}$ define the estimable parameters. In the sandwich variance estimator studied above, we consider $P=4$, $\hat{\boldsymbol{\eta}} = (\hat{\beta}_1 = \hat{\theta}(\delta), \hat{\beta}_2, \hat{\theta}_{00}=E_n[\frac{1-A}{P(A=1)} \frac{\hat{\pi}_A(\mathbf{X})}{1-\hat{\pi}_A(\mathbf{X})}(Y_1-Y_0-\hat{\mu}_{0,\Delta}(\mathbf{X}))], \hat{\theta}_{01}=E_n[\frac{A}{P(A=1)}\hat{\mu}_{0,\Delta}(\mathbf{X})])^T$.  Then,
\begin{align*}
    EE_{1i} &= \frac{A}{P(A=1)} \{ K(\frac{D_i-\delta}{h})(\hat{\xi}_i - \hat{\beta}_1 -  (\frac{D_i-\delta}{h}) \hat{\beta}_2) \\
    & \qquad + \int\limits_\mathbbm{D} K(\frac{D_i-\delta}{h}) (\hat{\mu}_{1,\Delta}(\mathbf{X_i}, d)-\hat{m}(d|A=1))d\hat{f}(d|A=1) \} \\
    EE_{2i} & = \frac{A}{P(A=1)} \{ K(\frac{D_i-\delta}{h})(\frac{D_i-\delta}{h})(\hat{\xi}_i - \hat{\beta}_1 -  (\frac{D_i-\delta}{h}) \hat{\beta}_2) \\
    & \qquad + \int\limits_\mathbbm{D} K(\frac{D_i-\delta}{h})(\frac{d-\delta}{h}) (\hat{\mu}_{1,\Delta}(\mathbf{X_i}, d)-\hat{m}(d|A=1))d\hat{f}(d|A=1) \} \\
    EE_{3i} &= (1-A_i) \{ \hat{\theta}_{00} - \frac{1}{P(A=1)} \frac{\hat{\pi}_A(\mathbf{X_i})}{1-\hat{\pi}_A(\mathbf{X_i})}(Y_{i1}-Y_{i0}-\hat{\mu}_{0,\Delta}(\mathbf{X_i})) \} \\
    EE_{4i} &= A_i \{ \hat{\theta}_{01} - \frac{1}{P(A=1)} \hat{\mu}_{0,\Delta}(\mathbf{X_i}) \} \\
\end{align*}
Then, since $\hat{\Psi}(\delta) = \hat{\beta}_1 - \hat{\theta}_{00} - \hat{\theta}_{01}$, we have $\hat{V}(\hat{\Psi}(\delta)) = V(1,1) + V(3,3) + V(4,4) + 2(V(3,4) - V(1,3) - V(1,4))$. 

Notably, $\boldsymbol{\eta}$ are not specific to the choice in nuisance function models, making this approach broadly applicable between studies. 
Still, several works have noted the benefits of including nuisance function uncertainty into the sandwich estimator in finite samples \citep{Stefanski2002TheM-Estimation, Cole2023IllustrationEstimators}.
In our setting, this may not be as important given the dominance, at least asymptotically, of the local linear kernel regression technique employed relative to nuisance function estimation. 
However, the sandwich estimator can be extended for example, by setting $\hat{\boldsymbol{\eta}} = (\hat{\beta}_1, \hat{\beta}_2, \hat{\theta}_{00}, \hat{\theta}_{01}, \boldsymbol{\alpha_D}, \sigma, \boldsymbol{\lambda_1}, \boldsymbol{\alpha_A}, \boldsymbol{\lambda_0})^T$ where $\boldsymbol{\alpha_D}$ and $\sigma$ are the parameters of the linear regression model for $\hat{\pi}_D$, $\boldsymbol{\lambda_1}$ are the parameters of the linear regression model for $\hat{\mu}_{1,\Delta}$, $\boldsymbol{\alpha_A}$ are the parameters of the logistic regression model for $\hat{\pi}_A$, and $\boldsymbol{\lambda_0}$ are the parameters of the linear regression model for $\hat{\mu}_{0,\Delta}$. 
The corresponding estimating equation components under this model specification are:
\begin{align*}
    EE_{6i} &= A_i(\frac{(D_i - \boldsymbol{\alpha_D}'\mathbf{X_i}^{(\pi_D)})^2}{\sigma^2}-1)\\
    EE_{7i} &= A_i\mathbf{X_i}^{(\mu_{1,\Delta})}{}'(Y_{i1}-Y_{i0} - \boldsymbol{\lambda_1}'\mathbf{X_i}^{(\mu_{1,\Delta})}) \\
    EE_{8i} &= \mathbf{X_i}^{(\pi_A)}{}'(A_i - \frac{exp(\boldsymbol{\alpha_A}'\mathbf{X_i}^{(\pi_A)})}{1+exp(\boldsymbol{\alpha_A}'\mathbf{X_i}^{(\pi_A)})})\\
    EE_{9i} &= (1-A_i) \mathbf{X_i}^{(\mu_{0,\Delta})}{}'(Y_{i1}-Y_{i0} - \boldsymbol{\lambda_1}'\mathbf{X_i}^{(\mu_{0,\Delta})}) \\
\end{align*}

When multiple observation times are observed, we can adjust for the correlation between $m$-time specific estimates with the sandwich estimator without specifying a particular correlation structure. To do so in the setting with $P=4$, we would set $\boldsymbol{\eta^*} = (\boldsymbol{\eta}^{(1)}{}^T, ..., \boldsymbol{\eta}^{(M)}{}^T)^T$ and $\Gamma^*_i = (\Gamma_{i1}^T, ..., \Gamma_{iM}^T)^T$ (i.e., stack each of the $M$ estimating equations specific to a given observation time) and then use the subsequent estimate of the covariance matrix $V$ to find $\hat{Var}(\frac{1}{M} \sum\limits_{m=1}^M \hat{\Psi}_m(\delta))$, while accounting for correlations across time like $\hat{Cov}(\hat{\Psi}_{m_1}(\delta), \hat{\Psi}_{m_2}(\delta))$.

\begin{table}
\caption{Comparison of Bias (RMSE) of effect curve estimates under multiply robust (MR), outcome regression (OR), and inverse probability weighting (IPW) approaches in simulation scenarios varying by nuisance function specification ($n=200$). Green cells represent scenarios where both sources of confounding are correctly adjusted for asymptotically, whereas orange and red cells represent scenarios where one and both sources are not correctly adjusted for, respectively. Confounding-naive and two-way fixed effects estimators have Bias (RMSE) of 0.323 (0.120) and 0.416 (0.255) in all scenarios.}
\label{t:sims_small}
\begin{center}
\begin{tabular}{lccc}
\hline
Incorrect Models & MR & OR & IPW \\
\hline
None & \cellcolor{green!40} 0.029 (0.048) & \cellcolor{green!40} 0.003 (0.043) & \cellcolor{green!40} 0.027 (0.092) \\
$\pi_A$ & \cellcolor{green!40} 0.028 (0.049) & \cellcolor{green!40} 0.003 (0.043) & \cellcolor{orange!40} 0.120 (0.124) \\
$\mu_{0,\Delta}$ & \cellcolor{green!40} 0.015 (0.056) & \cellcolor{orange!40} 0.082 (0.074) & \cellcolor{green!40} 0.027 (0.092) \\
$\pi_D$ & \cellcolor{green!40} 0.029 (0.048) & \cellcolor{green!40} 0.003 (0.043) & \cellcolor{orange!40} 0.122 (0.120) \\
$\mu_{1,\Delta}$ & \cellcolor{green!40} 0.043 (0.081) & \cellcolor{orange!40} 0.173 (0.091) & \cellcolor{green!40} 0.027 (0.092) \\
$\pi_A, \pi_D$ & \cellcolor{green!40} 0.028 (0.049) & \cellcolor{green!40} 0.003 (0.043) & \cellcolor{red!40} 0.158 (0.155) \\
$\mu_{0,\Delta}, \mu_{1,\Delta}$ & \cellcolor{green!40} 0.036 (0.089) & \cellcolor{red!40} 0.214 (0.135) & \cellcolor{green!40} 0.027 (0.092) \\
$\mu_{0,\Delta}, \pi_D$ & \cellcolor{green!40} 0.014 (0.056) & \cellcolor{orange!40} 0.082 (0.074) & \cellcolor{orange!40} 0.122 (0.120) \\
$\pi_A, \mu_{1,\Delta}$ & \cellcolor{green!40} 0.043 (0.082) & \cellcolor{orange!40} 0.173 (0.091) & \cellcolor{orange!40} 0.120 (0.124) \\
$\pi_A, \mu_{0,\Delta}$ & \cellcolor{orange!40} 0.133 (0.087) & \cellcolor{orange!40} 0.082 (0.074) & \cellcolor{orange!40} 0.120 (0.124) \\
$\pi_D, \mu_{1, \Delta}$ & \cellcolor{orange!40} 0.133 (0.107) & \cellcolor{orange!40} 0.173 (0.091) & \cellcolor{orange!40} 0.122 (0.120) \\
$\pi_A, \mu_{0,\Delta}, \pi_D$ & \cellcolor{orange!40} 0.133 (0.087) & \cellcolor{orange!40} 0.082 (0.074) & \cellcolor{red!40} 0.158 (0.155) \\
$\pi_A, \mu_{0,\Delta}, \mu_{1,\Delta}$ & \cellcolor{orange!40} 0.130 (0.118) & \cellcolor{red!40} 0.214 (0.135) & \cellcolor{orange!40} 0.120 (0.124) \\
$\pi_A, \pi_D, \mu_{1, \Delta}$ & \cellcolor{orange!40} 0.133 (0.107) & \cellcolor{orange!40} 0.173 (0.091) & \cellcolor{red!40} 0.158 (0.155) \\
$\mu_{0, \Delta}, \pi_D, \mu_{1, \Delta}$ & \cellcolor{orange!40} 0.129 (0.113) & \cellcolor{red!40} 0.214 (0.135) & \cellcolor{orange!40} 0.122 (0.120) \\
$\pi_A, \mu_{0, \Delta}, \pi_D, \mu_{1, \Delta}$ & \cellcolor{red!40} 0.175 (0.148) & \cellcolor{red!40} 0.214 (0.135) & \cellcolor{red!40} 0.158 (0.155) \\
\hline
\end{tabular}
\end{center}
\end{table}

\begin{table}
\caption{Comparison of Bias (RMSE) of effect curve estimates under multiply robust (MR), outcome regression (OR), and inverse probability weighting (IPW) approaches in simulation scenarios varying by nuisance function specification ($n=5000$). Green cells represent scenarios where both sources of confounding are correctly adjusted for asymptotically, whereas orange and red cells represent scenarios where one and both sources are not correctly adjusted for, respectively. Confounding-naive and two-way fixed effects estimators have Bias (RMSE) of 0.323 (0.120) and 0.417 (0.188) in all scenarios.}
\label{t:sims_large}
\begin{center}
\begin{tabular}{lccc}
\hline
Incorrect Models & MR & OR & IPW \\
\hline
None & \cellcolor{green!40} 0.021 (0.002) & \cellcolor{green!40} 0.002 (0.002) & \cellcolor{green!40} 0.029 (0.004) \\
$\pi_A$ & \cellcolor{green!40} 0.021 (0.002) & \cellcolor{green!40} 0.002 (0.002) & \cellcolor{orange!40} 0.129 (0.020) \\
$\mu_{0,\Delta}$ & \cellcolor{green!40} 0.008 (0.002) & \cellcolor{orange!40} 0.095 (0.011) & \cellcolor{green!40} 0.029 (0.004) \\
$\pi_D$ & \cellcolor{green!40} 0.020 (0.002) & \cellcolor{green!40} 0.002 (0.002) & \cellcolor{orange!40} 0.127 (0.027) \\
$\mu_{1,\Delta}$ & \cellcolor{green!40} 0.028 (0.004) & \cellcolor{orange!40} 0.182 (0.043) & \cellcolor{green!40} 0.029 (0.004) \\
$\pi_A, \pi_D$ & \cellcolor{green!40} 0.020 (0.002) & \cellcolor{green!40} 0.002 (0.002) & \cellcolor{red!40} 0.166 (0.047) \\
$\mu_{0,\Delta}, \mu_{1,\Delta}$ & \cellcolor{green!40} 0.013 (0.003) & \cellcolor{red!40} 0.235 (0.072) & \cellcolor{green!40} 0.029 (0.004) \\
$\mu_{0,\Delta}, \pi_D$ & \cellcolor{green!40} 0.007 (0.002) & \cellcolor{orange!40} 0.095 (0.011) & \cellcolor{orange!40} 0.127 (0.027) \\
$\pi_A, \mu_{1,\Delta}$ & \cellcolor{green!40} 0.028 (0.004) & \cellcolor{orange!40} 0.182 (0.043) & \cellcolor{orange!40} 0.129 (0.020) \\
$\pi_A, \mu_{0,\Delta}$ & \cellcolor{orange!40} 0.131 (0.020) & \cellcolor{orange!40} 0.095 (0.011) & \cellcolor{orange!40} 0.129 (0.020) \\
$\pi_D, \mu_{1, \Delta}$ & \cellcolor{orange!40} 0.126 (0.027) & \cellcolor{orange!40} 0.182 (0.043) & \cellcolor{orange!40} 0.127 (0.027) \\
$\pi_A, \mu_{0,\Delta}, \pi_D$ & \cellcolor{orange!40} 0.130 (0.020) & \cellcolor{orange!40} 0.095 (0.011) & \cellcolor{red!40} 0.166 (0.047) \\
$\pi_A, \mu_{0,\Delta}, \mu_{1,\Delta}$ & \cellcolor{orange!40} 0.139 (0.023) & \cellcolor{red!40} 0.235 (0.072) & \cellcolor{orange!40}  0.129 (0.020) \\
$\pi_A, \pi_D, \mu_{1, \Delta}$ & \cellcolor{orange!40} 0.126 (0.027) & \cellcolor{orange!40} 0.182 (0.043) & \cellcolor{red!40} 0.166 (0.047) \\
$\mu_{0, \Delta}, \pi_D, \mu_{1, \Delta}$ & \cellcolor{orange!40} 0.123 (0.025) & \cellcolor{red!40} 0.235 (0.072) & \cellcolor{orange!40} 0.127 (0.027) \\
$\pi_A, \mu_{0, \Delta}, \pi_D, \mu_{1, \Delta}$ & \cellcolor{red!40} 0.176 (0.051) & \cellcolor{red!40} 0.235 (0.072) & \cellcolor{red!40} 0.166 (0.047) \\
\hline
\end{tabular}
\end{center}
\end{table}

\begin{table}
\caption{Comparison of Bias (RMSE) of proposed estimate depending on $\xi$ regression specification ($n=1000$).}
\label{t:sims_kernel}
\begin{center}
\begin{tabular}{lcc}
\hline
Incorrect Models & Kernel & Parametric  \\
& & ($D+D^3$) \\
\hline
None & 0.027 (0.010) & 0.001 (0.008) \\
$\pi_A$ & 0.027 (0.010) & 0.001 (0.008) \\
$\mu_{0,\Delta}$ & 0.013 (0.011) & 0.022 (0.010) \\
$\pi_D$ & 0.027 (0.010) & 0.001 (0.008) \\
$\mu_{1,\Delta}$ & 0.029 (0.016) & 0.007 (0.012) \\
$\pi_A, \pi_D$ & 0.027 (0.010) & 0.001 (0.008) \\
$\mu_{0,\Delta}, \mu_{1,\Delta}$ & 0.016 (0.017) & 0.023 (0.014) \\
$\mu_{0,\Delta}, \pi_D$ & 0.013 (0.011) & 0.022 (0.010) \\
$\pi_A, \mu_{1,\Delta}$ & 0.029 (0.016) & 0.007 (0.012) \\
$\pi_A, \mu_{0,\Delta}$ & 0.139 (0.032) & 0.111 (0.024) \\
$\pi_D, \mu_{1, \Delta}$ & 0.128 (0.039) & 0.106 (0.029) \\
$\pi_A, \mu_{0,\Delta}, \pi_D$ & 0.139 (0.032) & 0.111 (0.024) \\
$\pi_A, \mu_{0,\Delta}, \mu_{1,\Delta}$ & 0.139 (0.038) & 0.110 (0.027) \\
$\pi_A, \pi_D, \mu_{1, \Delta}$ & 0.128 (0.039) & 0.106 (0.030) \\
$\mu_{0, \Delta}, \pi_D, \mu_{1, \Delta}$ & 0.124 (0.038) & 0.107 (0.031) \\
$\pi_A, \mu_{0, \Delta}, \pi_D, \mu_{1, \Delta}$ & 0.178 (0.066) & 0.145 (0.049) \\
\hline
\end{tabular}
\end{center}
\end{table}

\end{document}